\documentclass[ALICE,manyauthors]{cernphprep}

\pdfoutput=1
\usepackage[english]{babel}
\usepackage{graphicx}
\usepackage{verbatim}
\usepackage{amsfonts}
\usepackage{makeidx}
\usepackage{color}
\usepackage{amsmath}
\usepackage{lineno}
\usepackage{epsfig}
\usepackage{epstopdf}
\usepackage{hyperref}
\usepackage{cite}
\newcommand{\sqrts}{\sqrt{s}}

\newcommand{\GeV}{\mathrm{GeV}}
\newcommand{\TeV}{\mathrm{TeV}}

\newcommand{\gev}{\mathrm{GeV}}
\newcommand{\gevc}{\mathrm{GeV}/c}
\newcommand{\mevc}{\mathrm{MeV}/c}

\newcommand{\tev}{\mathrm{TeV}}

\newcommand{\mum}{\mathrm{\mu m}}

\newcommand{\mub}{\mathrm{\mu b}}

\newcommand{\pt}{p_{\rm T}}
\newcommand{\kt}{k_{\rm T}}

\newcommand{\DtoKpi}{{\rm D}^0 \to {\rm K}^-\pi^+}
\newcommand{\DtoKpipi}{{\rm D}^+\to {\rm K}^-\pi^+\pi^+}
\newcommand{\DstartoDpi}{{\rm D}^{*+} \to {\rm D}^0 \pi^+}
\newcommand{\DstartoDpiPrecise}{{\rm D^{*+}(2010)\to D^0\pi^+}}
\newcommand{\Dstophipi}{{\rm D_s^{+}\to \phi\pi^+}}

\newcommand{\DstophipitoKKpi}{{\rm D_s^{+}\to \phi\pi^+\to K^-K^+\pi^+}}
\newcommand{\DstoKKpi}{{\rm D_s^{+} \to K^-K^+\pi^+}}
\newcommand{\DplustoKpipi}{{\rm D}^+\to {\rm K}^-\pi^+\pi^+}

\newcommand{\DzerotoKpi}{{\rm D}^0 \to {\rm K}^-\pi^+}
\newcommand{\phitoKK}{{\rm \phi\to  K^-K^+}}

\newcommand{\Dzero}{{\rm D^0}}

\newcommand{\Dstar}{{\rm D^{*+}}}

\newcommand{\Dplus}{{\rm D^+}}

\newcommand{\Ds}{{\rm D_s^+}}

\newcommand{\meanpt}{{\langle p_{ {\mathrm T} } \rangle}}

\begin{document}

\begin{titlepage}
\PHyear{2017}
\PHnumber{020}      
\PHdate{30 January}  
%

\title{Measurement of D-meson production at mid-rapidity in pp collisions \\ at $\mathbf{\sqrts=7}$~TeV}
\ShortTitle{D-meson production in pp collisions at $\sqrts=7$~TeV}

\Collaboration{ALICE Collaboration\thanks{See Appendix~\ref{app:collab} for the list of collaboration members}}
\ShortAuthor{ALICE Collaboration} 

\begin{abstract} 
The production cross sections for prompt charmed 
mesons $\Dzero$, $\Dplus$, $\Dstar$ and $\Ds$
were measured at mid-rapidity in proton--proton collisions at a centre-of-mass energy 
$\sqrts=7~\tev$ with the ALICE detector at the Large Hadron Collider (LHC).
D mesons were reconstructed from their decays $\DtoKpi$, $\DtoKpipi$, 
$\DstartoDpi$, $\DstophipitoKKpi$, and their charge conjugates.
 With respect to previous measurements in the same rapidity region, the coverage in transverse momentum ($\pt$) is extended and the uncertainties are reduced by a factor of about two.  
The accuracy on the estimated total $\rm c\overline c$ production cross section is likewise improved.
The measured $\pt$-differential cross sections are compared with the results of three perturbative QCD calculations.
\end{abstract}

\end{titlepage}

\setcounter{page}{2}

\section{Introduction}
\label{sec:intro}


In high-energy hadronic collisions heavy quarks are produced by hard scatterings between partons of the two incoming hadrons. 
The production cross section of hadrons with charm or beauty quarks is calculated in the framework of Quantum Chromodynamics (QCD) and factorised as a convolution of the hard scattering cross sections at partonic level, the parton distribution functions (PDFs) of the incoming hadrons and the non-perturbative fragmentation functions of heavy quarks to heavy-flavour hadrons. 
Factorisation is implemented in terms of the squared momentum transfer $Q^2$ (collinear factorisation)~\cite{Collins:1989gx} or of the partonic transverse momentum $\kt$~\cite{Catani:1990eg}.
The hard scattering cross section is expanded in a perturbative series in powers of the strong coupling constant $\alpha_{\rm s}$.
 State-of-the-art calculations based on collinear factorisation implement a perturbative expansion up to next-to-leading order (NLO) in $\alpha_{\rm s}$, such as the general-mass variable flavour number scheme (GM-VFNS)~\cite{Kniehl:2004fy,Kniehl:2005mk,Kniehl:2012ti}, or next-to-leading order in $\alpha_{\rm s}$ with all-order resummation of the logarithms of $\pt/m_Q$  (FONLL)~\cite{Cacciari:1998it,Cacciari:2012ny}, where $\pt$ and $m_Q$ are the heavy-quark transverse momentum and mass, respectively. 
 Calculations based on $\kt$ factorisation exist only at leading order (LO) in $\alpha_{\rm s}$~\cite{Luszczak:2008je,Maciula:2013wg,Catani:1990eg}. All these calculations provide a good description of the production cross sections of 
 D and B mesons in proton--proton (and proton--antiproton) collisions at centre-of-mass energies from 0.2 to 13~TeV over a wide $\pt$ range at both central and forward rapidities (see e.g.~\cite{Andronic:2015wma} and references therein). In the case of charm production the uncertainties of the theoretical calculations, dominated by the perturbative scale uncertainties, are significantly larger than the experimental ones~\cite{Adamczyk:2012af,Acosta:2003ax,Aad:2015zix,ALICE:2011aa,Abelev:2012vra,Abelev:2012tca,Adam:2016ich,CMS-PAS-HIN-16-001,Aaij:2013mga,Aaij:2016jht,Aaij:2015bpa}. However, it was recently pointed out that in ratios of cross sections at different LHC energies and in different rapidity intervals the perturbative uncertainty becomes subdominant with respect to the uncertainty on the PDFs~\cite{Cacciari:2015fta}, thus making the measurement sensitive in particular to 
 the gluon PDF at values of Bjorken-$x$ down to $10^{-5}$ when the D-meson $\pt$ approaches 0. This represents a strong motivation for pursuing precise measurements of D-meson production in pp collisions at LHC energies.
Charm hadroproduction measurements are also required for cosmic-ray and neutrino astrophysics, where 
high-energy neutrinos from the decay of charmed hadrons produced in particle showers
in the atmosphere constitute
an important background for neutrinos from astrophysical 
sources~\cite{Bhattacharya:2015jpa,Bhattacharya:2016jce,Gauld:2015yia,Garzelli:2015psa}. 
 
In the context of the heavy-ion programme at the LHC, D-meson measurements in pp collisions represent an essential reference for the study of effects induced by cold and hot strongly-interacting matter
 in the case of proton--nucleus and nucleus--nucleus collisions (see e.g. the recent reviews~\cite{Andronic:2015wma,Prino:2016cni}).  
 In addition, the $\rm c\overline c$ production cross section per nucleon--nucleon collision
 is a basic ingredient for the determination of the amount of charmonium production by (re)generation in a quark-gluon plasma~\cite{Andronic:2011yq,Zhao:2011cv,Liu:2009nb}, 
 a mechanism that is supported by $\rm J/\psi$ measurements in nucleus--nucleus collisions at the LHC~\cite{Abelev:2013ila,Adam:2015isa}. A precise measurement of the $\rm c\overline c$ production cross section in pp collisions would enable a more stringent comparison of model calculations with data.
 
In this article, we report the measurement of the production cross sections of prompt $\rm D^0$, $\rm D^+$, $\rm D^{*+}$ and $\rm D_s^+$ mesons (as average of particles and anti-particles), and of their ratios,
in pp collisions at the centre-of-mass energy $\sqrt s=7~\tev$ using the ALICE detector at the LHC. The measurements cover  mid-rapidity ($|y|<0.5$) and 
the intervals $0<\pt<36~\gev/c$ for $\rm D^0$ mesons, $1<\pt<24~\gev/c$ for $\rm D^+$ and $\rm D^{*+}$mesons, and $2<\pt<12~\gev/c$ for $\rm D^+_s$ mesons.
The measurements cover complementary intervals in $\pt$ and rapidity with respect to those published by the ATLAS
($3.5<\pt<100~\gev/c$, $|\eta|<2.1$~\cite{Aad:2015zix}) and LHCb ($0<\pt<8~\gev/c$, $2<y<4.5$~\cite{Aaij:2013mga}) Collaborations
at the same centre-of-mass energy.
In comparison to previous ALICE publications based on the same data sample~\cite{ALICE:2011aa,Abelev:2012tca,Adam:2016ich}, the present results have a significantly extended $\pt$ coverage (for example, 
the previous coverage for $\rm D^0$ mesons was 0--16~$\gev/c$) and total uncertainties reduced by a factor of about two. These improvements have several sources: 
i) changes in the detector calibration, alignment and track reconstruction algorithm, which resulted in better $\pt$ resolution, thus higher signal-to-background ratio; ii) optimization of the D-meson selection procedure; iii) refinements in the estimation of the systematic uncertainties, which is now more data-driven; iv) a data sample with 20\% larger integrated luminosity. 
 
The article is organised as follows: the data sample and the analysis procedure are described in Section~\ref{sec:analysis}, the estimation of the systematic uncertainties is discussed in Section~\ref{sec:syst} and the results are presented and compared to theoretical calculations in Section~\ref{sec:results}.

\section{Analysis}
\label{sec:analysis}

A complete description of the ALICE experimental setup and of its performance
can be found in~\cite{Aamodt:2008zz,Abelev:2014ffa}. 
D mesons were reconstructed at mid-rapidity from their decay products,
using the tracking and particle identification capabilities of the 
ALICE central barrel detectors located within a 
large solenoidal magnet, providing a field B = $0.5~\mathrm{T}$ parallel to 
the beam line ($z$ axis of the ALICE reference frame).
The innermost detector, the Inner Tracking System (ITS), is used to track charged particles 
within the pseudorapidity interval $|\eta| < 0.9$ as well as for primary and secondary vertex reconstruction. It consists of six cylindrical layers equipped with Silicon Pixel Detectors (SPD), 
Silicon Drift Detectors (SDD) and Silicon Strip Detectors (SSD) from inner to outer layers. 
The ITS provides a resolution on the track impact parameter $d_0$ to the primary vertex 
in the transverse plane ($r\varphi$) better 
than 75~$\mu$m for transverse momentum $\pt >1~\gevc$.
As compared to previous publications based on the same data 
sample~\cite{ALICE:2011aa,Abelev:2012tca}, the alignment of the ITS sensor 
modules was improved and
a new procedure for the calibration of the drift velocity and of the 
non-uniformities of the drift field in the SDD was used.
The Time Projection Chamber (TPC) provides track reconstruction as well as particle identification via the measurement 
of the specific ionisation energy loss d$E$/d$x$.
The Time-Of-Flight detector (TOF) extends the charged particle identification capabilities of the TPC 
via the measurement of the flight time of the particles from the interaction point. 
The event collision time is measured with the T0 detector, which consists of
two arrays of Cherenkov counters 
located at $+350$~cm and $-70$~cm along the beam line, or, for the events with sufficiently large multiplicity, 
it is estimated using the particle arrival times at the TOF~\cite{Adam:2016ilk}.
The V0 detector, used in the online trigger and offline event selection, 
consists of two arrays of 32
scintillators each, covering the pseudorapidity intervals $-3.7 < \eta < -1.7$ and $2.8 < \eta < 5.1$, 
placed around the beam vacuum tube on either side of the interaction
region. A minimum-bias (MB) trigger was used to collect the data sample, by requiring at least one hit in either of the 
V0 counters or in the SPD ($|\eta| < 2$).  
Events were selected off-line by using the timing information from the V0 
and the correlation between the number of hits and track segments 
in the SPD detector to remove background due to beam--gas interactions. 
Only events with a primary vertex reconstructed within $\pm10~\rm{cm}$ from 
the centre of the detector along the beam line were used for the analysis.
The analysed data sample consists of about 370 million MB events, corresponding 
to an 
integrated luminosity $L_{\rm int} = (6.0 \pm 0.2)$~nb$^{-1}$, collected during
the 2010 pp run at $\sqrts = 7$~TeV. 

D mesons were reconstructed via their hadronic 
decay channels $\DzerotoKpi$ (with branching ratio, BR = 3.93 $\pm$ 0.04\%), 
$\DplustoKpipi$ (BR = 9.46 $\pm$ 0.24\%), $\DstartoDpiPrecise$ 
(strong decay with BR = 67.7 $\pm$ 0.5\%) with $\DzerotoKpi$ and 
$\Dstophipi$ (BR = 2.27 $\pm$ 0.08\%) with $\phitoKK$, together with their 
charge conjugates~\cite{Olive:2016xmw}. 


D-meson candidates were defined using pairs or
triplets of tracks with the proper charge-sign combination.
Tracks were required to have $|\eta| < 0.8$, $\pt > 0.3~\gev/c$, 
at least 70 associated TPC space points (out of a maximum of 159), $\chi^{2}/{\rm ndf} < 2$ in 
the TPC (where ndf is the number of degrees of freedom involved in the 
track fit procedure), 
and at least one hit in either of the two layers of the SPD.  For the soft pion
produced in $\Dstar$ decay, also tracks reconstructed only with the ITS, with at least four hits, including 
at least one in the SPD, and $\pt >$ 80 $\mevc$ were considered.
With these track selection criteria, the acceptance in rapidity for D mesons drops steeply to zero for 
$|y|>0.5$ 
at low $\pt$ and $|y|>0.8$ at $\pt>5~\gevc$. A $\pt$-dependent fiducial acceptance cut was 
therefore applied on the D-meson rapidity, $|y| < y_{\rm fid}(\pt)$, with $y_{\mathrm{fid}}(\pt)$ increasing from 
0.5 to 0.8 in the transverse momentum range $0 < \pt < 5~\gevc$ according to a second-order polynomial function, 
and $y_{\mathrm{fid}}=0.8$ for $\pt > 5~\gevc$.

$\Dzero$, $\Dplus$ and $\Ds$ mesons have mean proper decay lengths $c\tau$ 
of about 123, 312 and 150~$\mu$m, respectively~\cite{Olive:2016xmw}. 
Their decay vertices are therefore typically displaced by a few hundred $\mu$m from the primary
vertex of the interaction. Geometrical selections on the D-meson decay topology were applied to 
reduce the combinatorial background. 
The selection requirements were tuned so as to provide a large statistical 
significance for the signal and to keep the selection efficiency as high as 
possible.
The latter requirement was dictated also by the fact that too tight cuts 
result in an increased contribution to the raw yield from feed-down D mesons 
originating from decays of B mesons. In the $\DstartoDpi$
case, the decay vertex cannot be resolved from the primary vertex and
geometrical selections were applied on the secondary vertex topology of the 
produced $\Dzero$.
The geometrical selections were mainly based on the 
displacement of the tracks from the interaction vertex, the distance between 
the D-meson decay vertex and the primary vertex (decay length, $L$), and the 
pointing of the reconstructed D-meson momentum to the primary vertex. 
The pointing condition is applied by requiring a small value for the 
angle $\theta_{\rm pointing}$ between the directions of
the reconstructed momentum of the candidate and its flight line, defined 
by the vector from the primary to the secondary vertex.
In comparison to the previous analysis of the same data sample,
additional selection criteria were introduced. 
In particular, the projections of the pointing angle and of the decay 
length in the transverse plane ($\theta_{\rm pointing}^{r\varphi}$ and $L^{r\varphi}$) 
were considered. Moreover, a cut on the normalised difference between the measured and expected impact parameters of each of the decay particles 
$(d^{\rm reco}_{0,tr} - d^{\rm exp}_{0,tr}$)/$\sigma_{\Delta}$ was applied, where 
$d^{\rm reco}_{0,tr}$ is the measured track impact parameter, 
$d^{\rm exp}_{0,tr}$ is defined as $L^{r\varphi} \sin(\theta_{\rm tr,D}^{r\varphi})$,
$\theta_{\rm tr,D}^{r\varphi}$ is the measured angle between the momenta 
of the D meson and of the considered track, and $\sigma_{\Delta}$ is the 
combination of the uncertainties on the measured and expected $d_{0}$.
By requiring $(d^{\rm reco}_{0,tr} - d^{\rm exp}_{0,tr})/\sigma_{\Delta}<3$, a significant rejection of 
background candidates (15--40\% depending on D-meson species and $\pt$) and feed-down D mesons (up to 50\% at high $\pt$) 
is achieved while keeping almost 100\% of the prompt D mesons.

Further reduction of the combinatorial background was obtained by applying particle identification (PID) to the decay tracks. 
A $3 \sigma$ compatibility cut was applied on the difference between the measured and expected signals for pions 
and kaons for both the $\mathrm{d}E/\mathrm{d}x$ and time-of-flight. 
Tracks without TOF hits were identified using only the TPC information with a $3\,\sigma$ selection
for $\Dzero$, $\Dplus$ and $\Dstar$ decay products, and a $2\,\sigma$ selection for the $\Ds$.
This stricter PID selection strategy was needed in the $\Ds$ case
due to the large background of track triplets and the
short $\Ds$ lifetime, which limits the effectiveness of the geometrical 
selections on the displaced decay-vertex topology.
Based on the PID information and the charge sign of the decay tracks,
$\Dzero$ candidates were accepted (as $\rm{D^{0}}$, $\rm{\overline{D^0}}$, or both) or rejected, according 
to the compatibility with the K$^{\mp}$$\pi^{\pm}$ final state. For the $\Dstar$ reconstruction, 
this ambiguity is resolved using the charge of the soft pion.
In the cases of the $\DstoKKpi$ and $\DplustoKpipi$ decays, a candidate was rejected if 
the track with charge opposite to that of the D meson was not compatible with the kaon PID hypothesis.

The D-meson raw yields, including both particles and anti-particles, were obtained from fits to the $\Dzero$, 
$\Dplus$ and $\Ds$ candidate invariant-mass distributions and to the mass difference 
$\Delta M = M (\mathrm{K} \pi \pi) - M(\mathrm{K} \pi)$ distributions for $\Dstar$ candidates. 
In the fit function, the signal was modeled with a Gaussian and the background was described by an 
exponential term for $\Dzero$, $\Dplus$ and $\Ds$ candidates and by the function 
$a \sqrt{\Delta M - m_{\pi}} \cdot {\rm e}^{b(\Delta M - m_{\pi})}$ for $\Dstar$ candidates. 
In the case of  $\Dzero$ mesons, an additional term was included in
the fit function to account for the contribution of signal candidates that are present in the invariant mass distribution 
with the wrong daughter particle mass assignment (reflections).  A study with Monte Carlo simulations 
showed that about 70$\%$ of these reflections are rejected by the PID 
selections. The residual contribution was accounted for by including in the fit a
template consisting of the sum of two wide Gaussians with centroids and widths fixed to values obtained in the simulation
 and with amplitudes normalised using the signal observed in data.

\begin{figure}[!thb]
\begin{center}
\includegraphics[width=1.\textwidth]{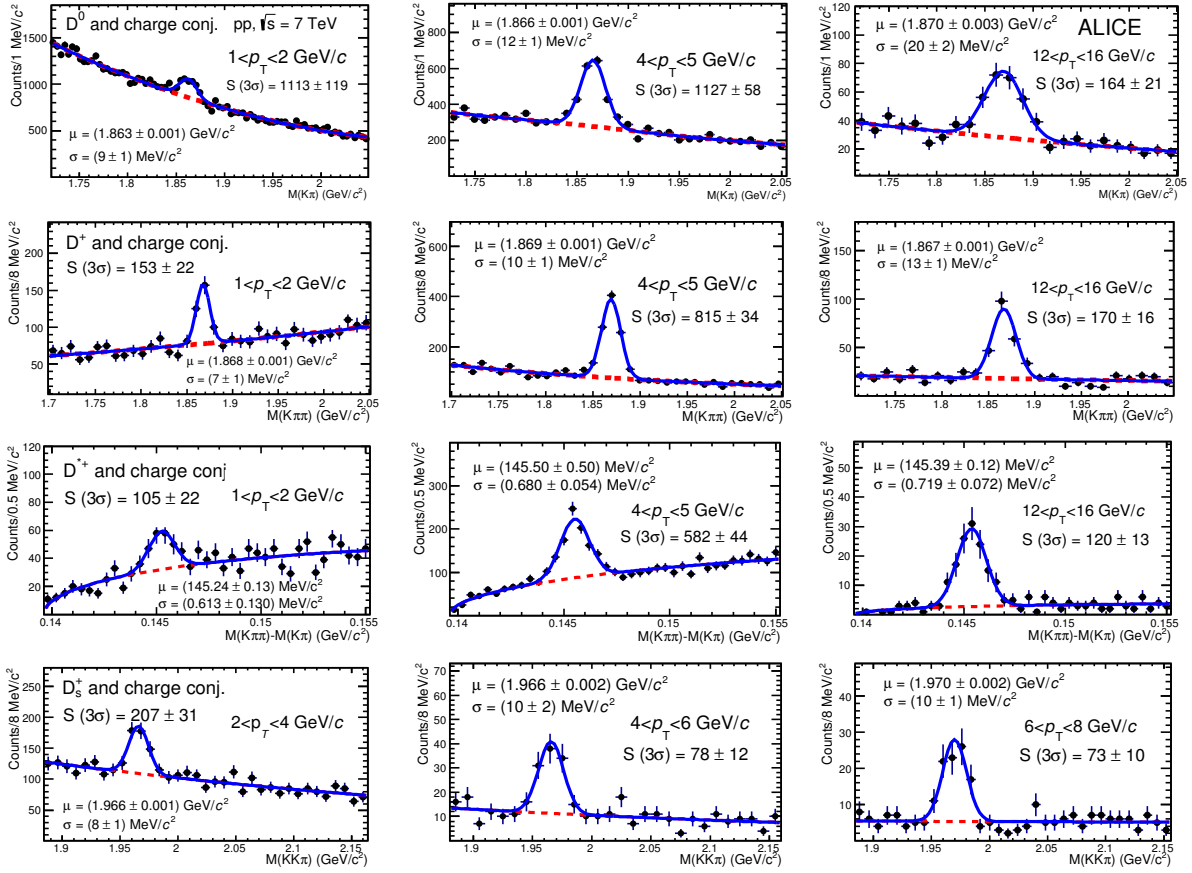}
\caption{Invariant-mass (mass-difference) distributions of $\Dzero$, $\Dplus$, ($\Dstar$) and $\Ds$ 
candidates and charge conjugates in three $\pt$ intervals for a sample of pp collisions at $\sqrt s=7~\tev$ 
with $L_{\rm int}=6.0~{\rm nb}^{-1}$. The curves show the fit functions as described in the text.
The contribution of reflections for the $\Dzero$ meson is included.  
The values of mean ($\mu$) and width ($\sigma$) of the signal peak are 
reported together with the signal counts ($S$) in the mass interval ($\mu$- 3$\sigma,\mu + 3\sigma$).}
\label{fig:InvMass}
\end{center}
\end{figure}

Figure~\ref{fig:InvMass} shows fits to the
invariant-mass (mass-difference) distributions in three $\pt$ intervals for $\Dzero$, $\Dplus$, ($\Dstar$) 
and $\Ds$ candidates from top to bottom. 
The mean values of the Gaussians in all transverse-momentum intervals 
were found to be compatible within uncertainties with the world average rest 
mass values for $\Dzero$, $\Dplus$ and $\Ds$ and with the difference 
$M_{\Dstar} - M_{\Dzero}$ for the $\Dstar$~\cite{Olive:2016xmw}.
The widths are consistent with the results from Monte Carlo simulations and smaller by 10--20\% 
than the values in~\cite{ALICE:2011aa,Abelev:2012tca}, as a consequence of 
the improved $\pt$ resolution.

The $\pt$-differential cross section of prompt D mesons was computed as:
\begin{equation}
  \label{eq:dNdpt}
  \frac{{\rm d}^2 \sigma^{\rm D}}{{\rm d}\pt {\rm d} y}=
  \frac{1}{c_{\Delta y} ~\Delta \pt}\frac{1}{{\rm BR}}  \frac{\frac{1}{2} \, \left.f_{\rm prompt}\cdot N^{\rm D+\overline D,raw}\right|_{|y|<y_{\rm fid}}}{ ({\rm Acc}\times\epsilon)_{\rm prompt}}    \frac{1}{L_{\rm int}} \,,
\end{equation}
where $f_{\rm prompt}$, $N^{\rm D+\overline D,raw}$ and $({\rm Acc}\times\epsilon)_{\rm prompt}$ are  $\pt$-interval dependent quantities. 
The raw yield values (sum of particles and antiparticles, 
$N^{\rm D+\overline D,raw}$) were corrected for the B-meson decay feed-down 
contribution (i.e.\,multiplied by the prompt fraction $f_{\rm{prompt}}$ in the 
raw yield), divided 
by the acceptance-times-efficiency for prompt D mesons
$(\rm Acc \times \epsilon)_{\rm{prompt}}$, and divided by a factor of two to obtain the 
particle and antiparticle averaged yields.
The $\pt$-differential yields for each D-meson species, measured separately 
for particles and anti-particles, were found to be in agreement within statistical uncertainties.
The corrected yields were divided by the decay channel BR, 
the $\pt$ interval width $\Delta \pt$, the correction factor for the
rapidity coverage $c_{\Delta y}$, and the integrated luminosity $L_{\rm int}=N_{\rm ev} / \sigma_{\rm MB} $, where $N_{\rm ev}$ 
is the number of analysed events and $\sigma_{\rm MB}=62.2$~mb is the cross section for the MB trigger condition \cite{Abelev:2012sea}.

\begin{figure}[!t]
\begin{center}
\includegraphics[width=1.\textwidth]{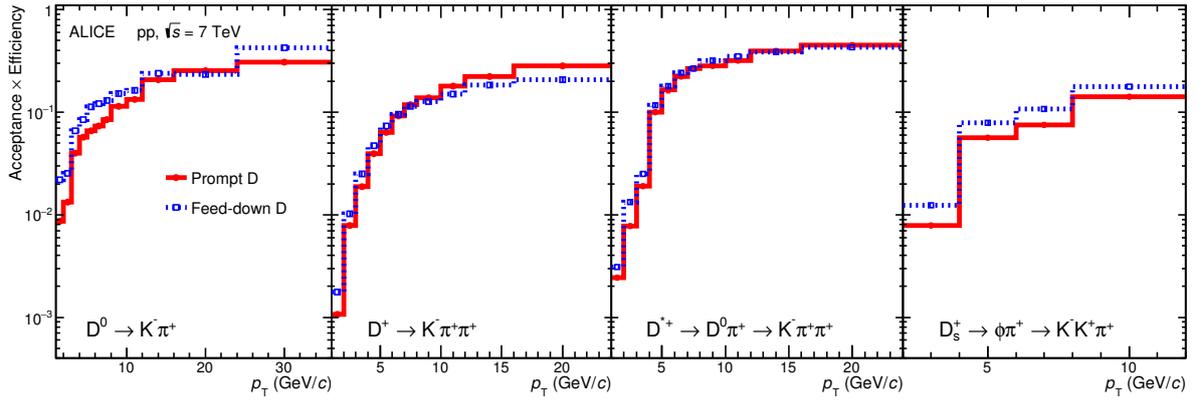}
\caption{Acceptance $\times$ efficiency for $\Dzero$, $\Dplus$, $\Dstar$ and $\Ds$ mesons, as a function of $\pt$.
The efficiencies for prompt (solid lines) and feed-down (dotted lines)
D mesons are shown.}
\label{fig:AccEff}
\end{center}
\end{figure}

The  $(\rm Acc \times \epsilon)$ correction factor was determined using simulations 
of pp collisions generated with the PYTHIA\,6.4.21 event generator~\cite{Sjostrand:2006za} (Perugia-0 tune~\cite{Skands:2010ak}), 
and particle transport through the apparatus 
using GEANT3~\cite{Brun:1994aa}.
The luminous region distribution and the conditions 
of all the ALICE detectors were included in the 
simulations.
The $(\rm Acc \times \epsilon)$ for prompt and feed-down $\Dzero$, $\Dplus$, 
$\Dstar$ and $\Ds$ mesons with $|y| < y_{\rm fid}$ is shown in Fig.~\ref{fig:AccEff} as a function of $\pt$.
The efficiencies for feed-down D mesons are higher than those for prompt D mesons in most of 
the $\pt$ intervals, because the decay vertices of the feed-down D mesons 
are on average more displaced from the primary vertex due to the large B-meson lifetime 
($c\tau\approx 500~\mum$~\cite{Olive:2016xmw}). 
However, the selection on the difference between measured and expected decay-track 
impact parameters rejects more efficiently feed-down D mesons, thus reducing the difference between 
prompt and feed-down efficiencies as compared to the previous analyses.

The rapidity acceptance correction factor $c_{\Delta y}$ was computed with
the PYTHIA 6.4.21 event generator with Perugia-0 tune as the ratio between the
generated D-meson yield in $\Delta y = 2\,y_{\rm fid}$,
(with $y_{\rm fid}$ varying from 0.5 at low $\pt$ to 0.8 at high $\pt$)
and that in $|y|<0.5$.
It was checked that calculations of the $c_{\Delta y}$ correction factor
based on FONLL pQCD calculations~\cite{Cacciari:2012ny} or
on the assumption of uniform D-meson rapidity distribution in $|y|<y_{\rm fid}$
would give the same result, because both in PYTHIA and in FONLL
the D-meson yield is uniform within 1\% in the range $|y|<0.8$.

The $f_{\rm prompt}$ fraction was calculated using the B production cross sections from  
FONLL calculations~\cite{Cacciari:1998it, Cacciari:2001td}, the 
$\mathrm{B} \rightarrow \mathrm{D} + X$ decay kinematics from the EvtGen package~\cite{Lange:2001uf} 
and the efficiencies for feed-down D mesons reported in 
Fig.~\ref{fig:AccEff}:
\begin{equation}
\label{eq:fpr}
f_{\mathrm{prompt}} =1- \frac{N^{\text{D~feed-down}}_{\mathrm{raw}}}{N^{\mathrm{D}}_{\mathrm{raw}}}= 1- \left (\frac{\rm d^2 \sigma}{\mathrm d\pt \mathrm d y} \right)^{\rm FONLL}_{\text{feed-down}} \cdot \frac{(\mathrm{Acc} \times \epsilon)_\text{feed-down} \cdot \Delta y \Delta \pt \cdot \mathrm{BR} \cdot L_{\rm int}}{N^{\rm D +\overline{D},raw}/2}\,,
\end{equation}
where the $\pt$ dependence of $f_{\rm prompt}$, $N^{\rm D +\overline{D},raw}$ and
$(\mathrm{Acc} \times \epsilon)_\text{feed-down}$ is omitted for brevity.
The values of $f_{\rm prompt}$ range between 0.85 and 0.97 depending on 
D-meson species and $\pt$.

\section{Systematic uncertainties}
\label{sec:syst}


Systematic uncertainties were estimated considering several sources. 
A summary is shown in Table~\ref{tab:syst}  for two $\pt$ intervals. 
New or refined procedures were used with respect to the analyses presented in~\cite{ALICE:2011aa,Abelev:2012tca}, in particular for the uncertainties on the signal yield extraction, the track reconstruction efficiency and the feed-down subtraction.

The systematic uncertainties on the yield extraction obtained from the fits to the invariant-mass distributions (mass difference for $\Dstar$ mesons) were evaluated by repeating the fits several times varying (i) the invariant-mass bin width, (ii) the lower and upper limits of the fit range, (iii) the background fit function (exponential function, first, second and third order polynomials were used for $\Dzero$, $\Dplus$ and $\Ds$ and a power law for the $\Dstar$), for a total of about few hundred fits for each D-meson species and $\pt$ interval.  
In addition, the same approach was used with a bin counting method, in which the signal yield was obtained by integrating the invariant-mass distribution 
after subtracting the background estimated from a fit to the side-bands.
The distributions of the signal yield obtained from these variations are consistent with a Gaussian shape and the mean of the distributions is close to the central value of the yield. The systematic uncertainty was defined as the R.M.S.\, of this distribution.

The systematic uncertainty on the track reconstruction efficiency was estimated by varying the track-quality selection criteria and by comparing the probability to prolong tracks from the TPC inward to the ITS (`matching efficiency') in data and simulations. The variation of the track selection criteria, such as
the minimum number of clusters in the TPC, was found to yield a 2\% systematic 
effect on the cross section of $\Dzero$ mesons (two-prong final state) and 3\% for the other meson species (three-prong final states). The comparison of the matching efficiency in data and simulations was made after weighting the relative abundances of primary and secondary particles in the simulation to match those observed in data. 
This weighting is motivated by the observation that the matching efficiency is much larger for primary particles than  for secondary particles produced far from the interaction point in decays of strange hadrons and in interactions of primary particles with the material of the detector. 
The fractions of primary and secondary particles were estimated, as a function of $\pt$, by fitting the inclusive track impact parameter distributions in data and in the simulation with a sum of three template distributions for primary particles, for secondary particles from strange-hadron decays and for secondary particles produced in interactions of primary particles in the detector material. The templates were obtained from the simulation.
After weighting the relative abundances in the simulation to match those in data, the systematic uncertainty on the matching efficiency was defined as the relative difference of the matching efficiencies 
in data and in the simulation. The study was made separately for particles identified as pions and as kaons using the TPC and TOF PID selections described in Section~\ref{sec:analysis}.
The systematic uncertainty is 2\% per track in the interval  $2<\pt<6$~GeV/$c$ and 1\% at lower and higher $\pt$.
The per-track uncertainty was then propagated to the D mesons, taking into account the number and transverse momentum of their decay tracks,
and added in quadrature to the component estimated from the track selection variation.

Systematic uncertainties can also arise from possible differences in the distributions and resolution of the geometric selection variables between data and the simulation.
These uncertainties were evaluated by repeating the analysis with several sets of selection criteria and comparing the resulting corrected cross sections.
More details can be found in \cite{ALICE:2011aa}.

\begin{table}[!t]
\centering
\begin{tabular}{l|cc|cc|cc|cc}
\hline
 & \multicolumn{2}{c|}{$\Dzero$}
 & \multicolumn{2}{c|}{$\Dplus$}
 & \multicolumn{2}{c|}{$\Dstar$}
 & \multicolumn{2}{c}{$\Ds$}\\
 $\pt$ (GeV/$c$) & 2--3   & 10--12 
 & 2--3   & 10--12 
 & 2--3   & 10--12 
 & 2--4   & 8--12  \\
\hline
Signal yield  & 3\% & 4\%
                      & 6\% & 5\%
                      & 2\% & 2\%
                      & 5\% & 5\% \\
Tracking efficiency  & {4\%} & {4\%}  
                      & {4\%} & {6\%}
                      & {6\%} & {6\%}
                      & {5\%} & {6\%} \\
Selection efficiency        & {5\%} & {5\%}
                      & 10\% & 5\%
                      & 5\% & 5\%
                      & {7\%} & {7\%}\\
PID efficiency        & {0} & {0}
                      & {0} & {0}
                      & {0} & {0}
                      & {7\%} & {7\%}\\
$\pt$ shape in MC        & {0} & {0}
                      & {1\%} & {2\%}
                      & {2\%} & {0}
                      & { 3\%} & {2\%}\\
Feed-down      & $^{+4}_{- 4}\%$ & $^{+3}_{- 5}\%$
                  & $^{+2}_{- 3}\%$ & $^{+2}_{- 3}\%$
                  & $^{+2}_{- 2}\%$ & $^{+2}_{- 3}\%$
                  & $^{+4}_{- 5}\%$ & $^{+4}_{- 5}\%$\\
Branching ratio       
                      & \multicolumn{2}{c|}{1.0\%}
                      & \multicolumn{2}{c|}{2.5\%}
                      & \multicolumn{2}{c|}{1.3\%}
                      & \multicolumn{2}{c}{3.5\%} \\


\hline
Normalisation         & \multicolumn{8}{c}{3.5\%}\\
\hline
\end{tabular}
\caption{Summary of relative systematic uncertainties for two $\pt$ intervals.}
\label{tab:syst}
\end{table}

To estimate the uncertainty on the PID selection efficiency, for the three non-strange D-meson species the analysis was repeated without PID selection. The resulting cross sections were found to be compatible with those obtained with the PID selection. 
Therefore, no systematic uncertainty was assigned.
For the $\Ds$ meson, the lower signal yield and the larger combinatorial background prevented a signal estimation without
particle identification, hence, in this case, a  3$\sigma$ PID selection, looser with respect to the PID strategy adopted in the analysis, was used to estimate a systematic uncertainty of about 7\%.
 
 The systematic effect on the efficiency due to a possible difference between the simulated and real $\pt$ distribution of D mesons was estimated by using alternative D-meson $\pt$ distributions from the PYTHIA\,6 generator with Perugia-0 tune
and from the FONLL pQCD calculation. More details can be found in \cite{ALICE:2011aa}. 

The systematic uncertainty on the subtraction of feed-down from beauty-hadron decays includes the uncertainties of i) the B-meson
production cross section from FONLL calculations, ii) the branching ratios of B mesons into D mesons~\cite{Olive:2016xmw} and iii) the relative abundances
of B-meson species produced in the beauty-quark fragmentation~\cite{Olive:2016xmw}.
The dominant contribution is the one originating from the FONLL
calculations and it was estimated by varying the $\pt$-differential cross section of feed-down D mesons 
within the theoretical uncertainties of the FONLL calculation. The procedure for the variation of the b-quark mass, of the perturbative scales and of the parton distribution functions is 
described in~\cite{Cacciari:2012ny}.
In previous analyses, an alternative method based on the ratio of the FONLL cross sections for feed-down and prompt D mesons was also used in the estimation of the systematic uncertainties. In this analysis it  was no longer used, on the basis of the observation that FONLL calculations at LHC energies provide a good description of the production cross sections of  $\rm B^{0}$, $\rm B^{+}$ and $\rm B_s^0$ mesons at both central and forward rapidity, while it underestimates prompt charm production ~\cite{Aaij:2013noa,Chatrchyan:2011vh,Khachatryan:2011mk,Chatrchyan:2011pw,Abelev:2012sca,Abelev:2012gx}. 
Hence, the uncertainty due to the B feed-down correction is significantly 
reduced and more symmetric as compared to our previous publications.
 
The uncertainty on the D-meson production cross section normalisation has a contribution from the 3.5\% uncertainty on the minimum-bias trigger cross section  \cite{Abelev:2012sea} and a contribution from the uncertainties on the branching ratios of the considered D-meson decay channels (see Table~\ref{tab:syst}).

The total systematic uncertainties, which are obtained as a quadratic sum of the contributions listed in Table~\ref{tab:syst}, are reduced by a factor that ranges from 1.5 to 5,
depending on D-meson species and $\pt$ interval, with respect to previous publications~\cite{ALICE:2011aa,Abelev:2012tca}. 
The systematic uncertainties on PID, tracking and selection 
efficiencies are mostly correlated among the different $\pt$ intervals, 
while the raw-yield extraction uncertainty is mostly uncorrelated.

\section{Results}
\label{sec:results}

\begin{figure}[!t]
\begin{center}
\includegraphics[width=.6\textwidth]{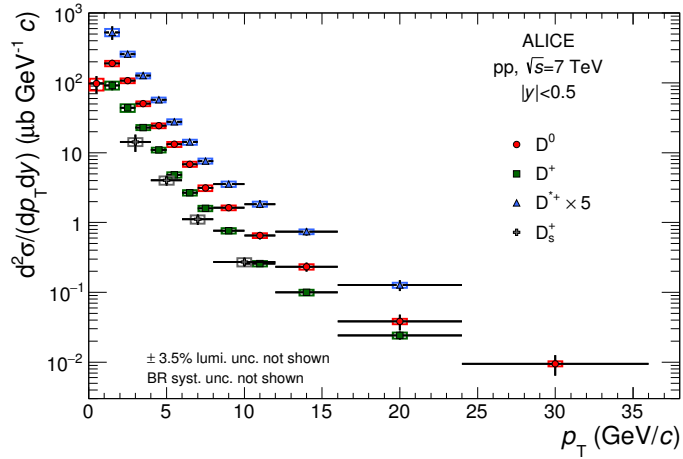}
\caption{$\pt$-differential inclusive production cross section of prompt
$\Dzero$, $\Dplus$, $\Dstar$ and $\Ds$ mesons in pp collisions
at $\sqrts=7~\tev$. Statistical uncertainties (bars) and systematic 
uncertainties (boxes) are shown. The $\Dstar$ cross section is scaled by a 
factor of 5 for better visibility.}
\label{fig:crosssec4} 
\end{center}
\end{figure}

The $\pt$-differential cross sections for prompt $\Dzero$, $\Dplus$, $\Dstar$ 
and $\Ds$ production in $|y|<0.5$ are shown in Fig.~\ref{fig:crosssec4}.
The error bars represent the statistical uncertainties, while the systematic 
uncertainties are shown as boxes around the data points. 
The symbols are positioned horizontally at the centre of each $\pt$ interval,
with the horizontal bars representing the width of the $\pt$ interval. 
For all D-meson species, the results are consistent within 
uncertainties with those reported in our previous publications on charmed-meson 
cross sections in pp collisions at 
$\sqrts=7~\tev$~\cite{ALICE:2011aa,Abelev:2012tca},
but the total uncertainties (sum in quadrature of statistical and systematic
errors) are reduced by a factor 1.5--4, 
depending on the D-meson species and the $\pt$ interval.
The $\Dzero$-meson cross section in the interval $0<\pt<1~\gev/c$ is obtained 
from the analysis without decay vertex reconstruction described in 
Ref.~\cite{Adam:2016ich}.
At higher $\pt$, the results of the analysis presented in this paper, based on
geometrical selections on the displaced decay vertex, are more precise than 
those obtained without decay vertex reconstruction.

\begin{figure}[!thb]
\begin{center}
\includegraphics[width=0.48\textwidth]{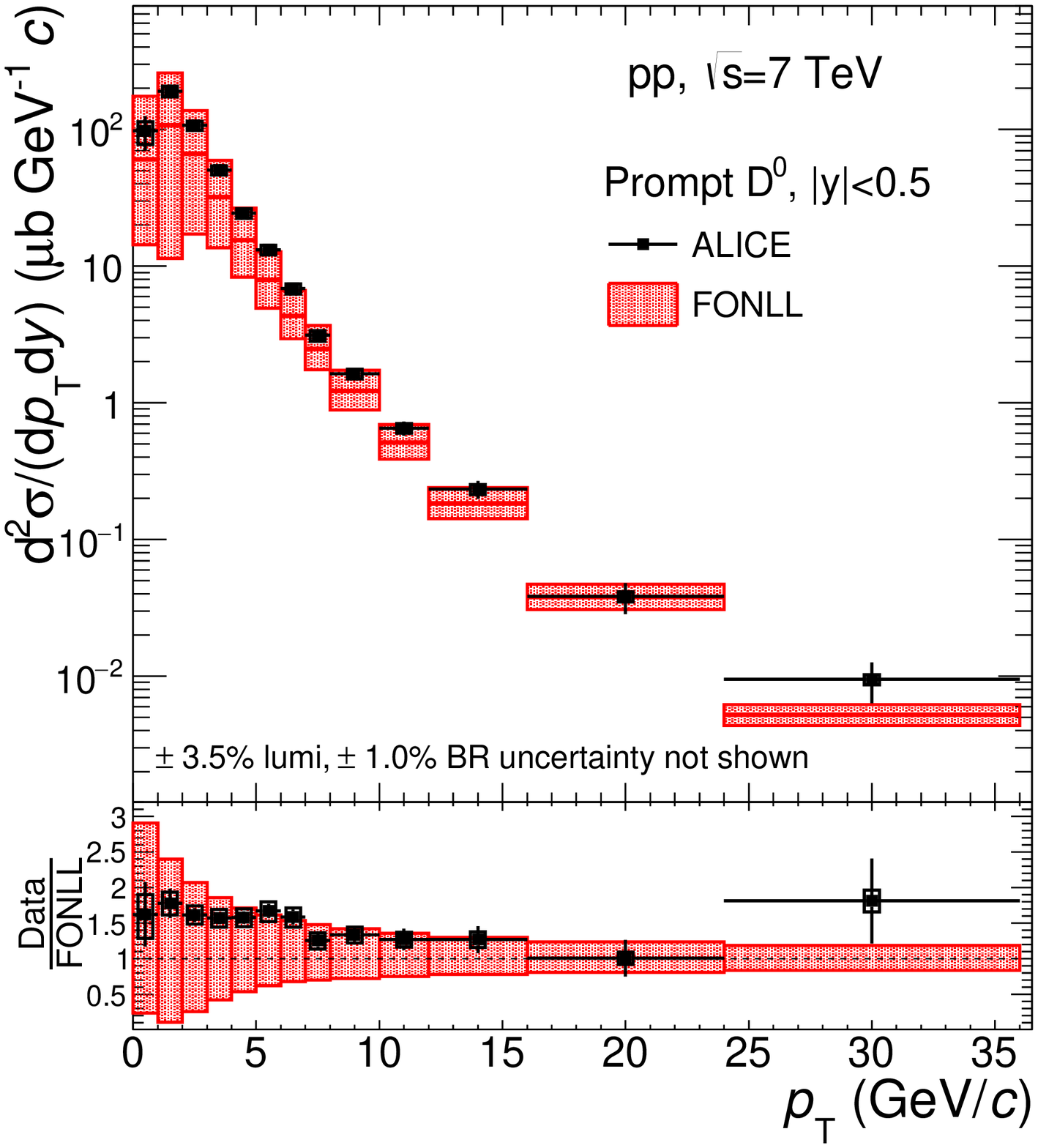}
\includegraphics[width=0.48\textwidth]{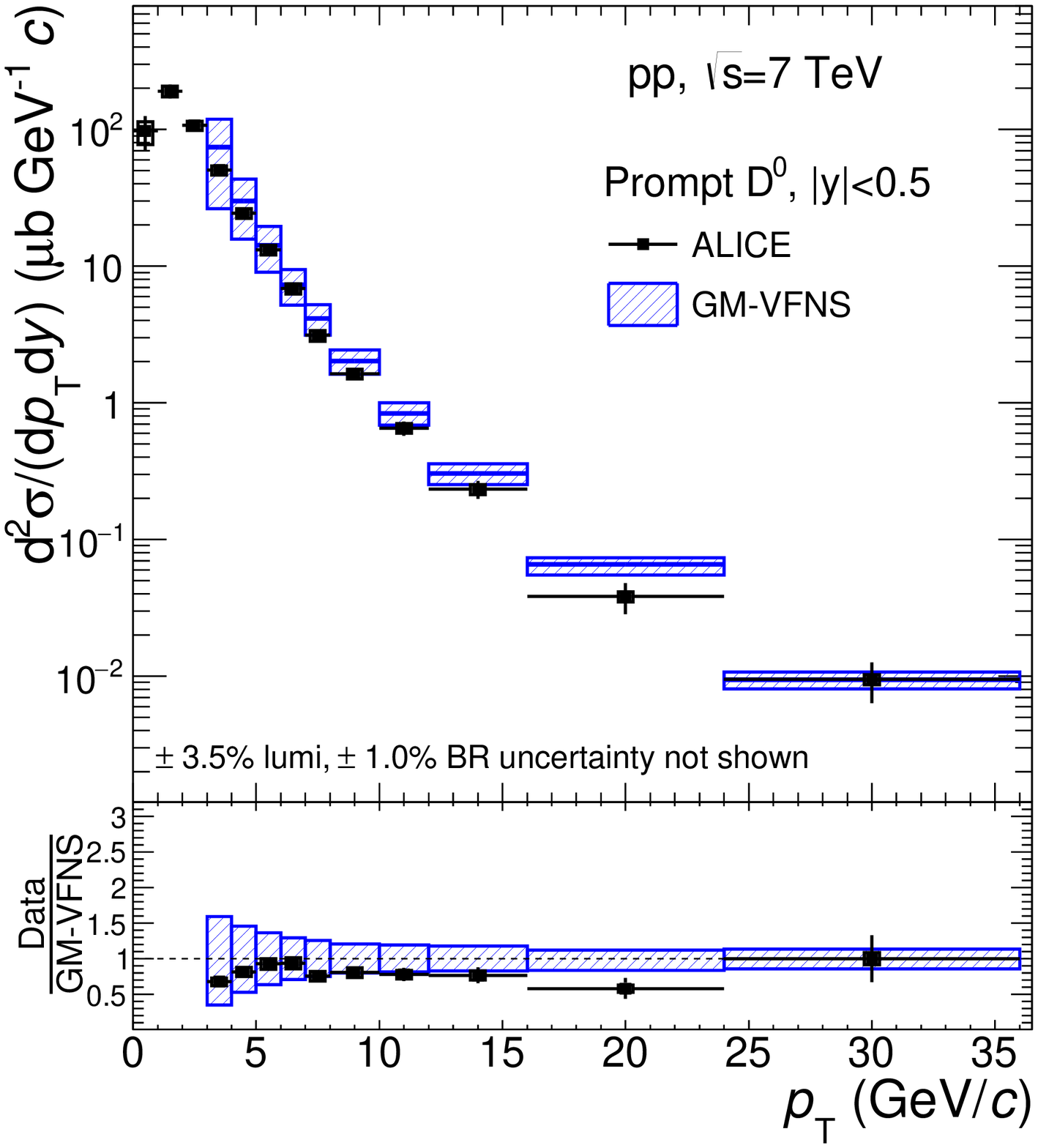}
\includegraphics[width=0.48\textwidth]{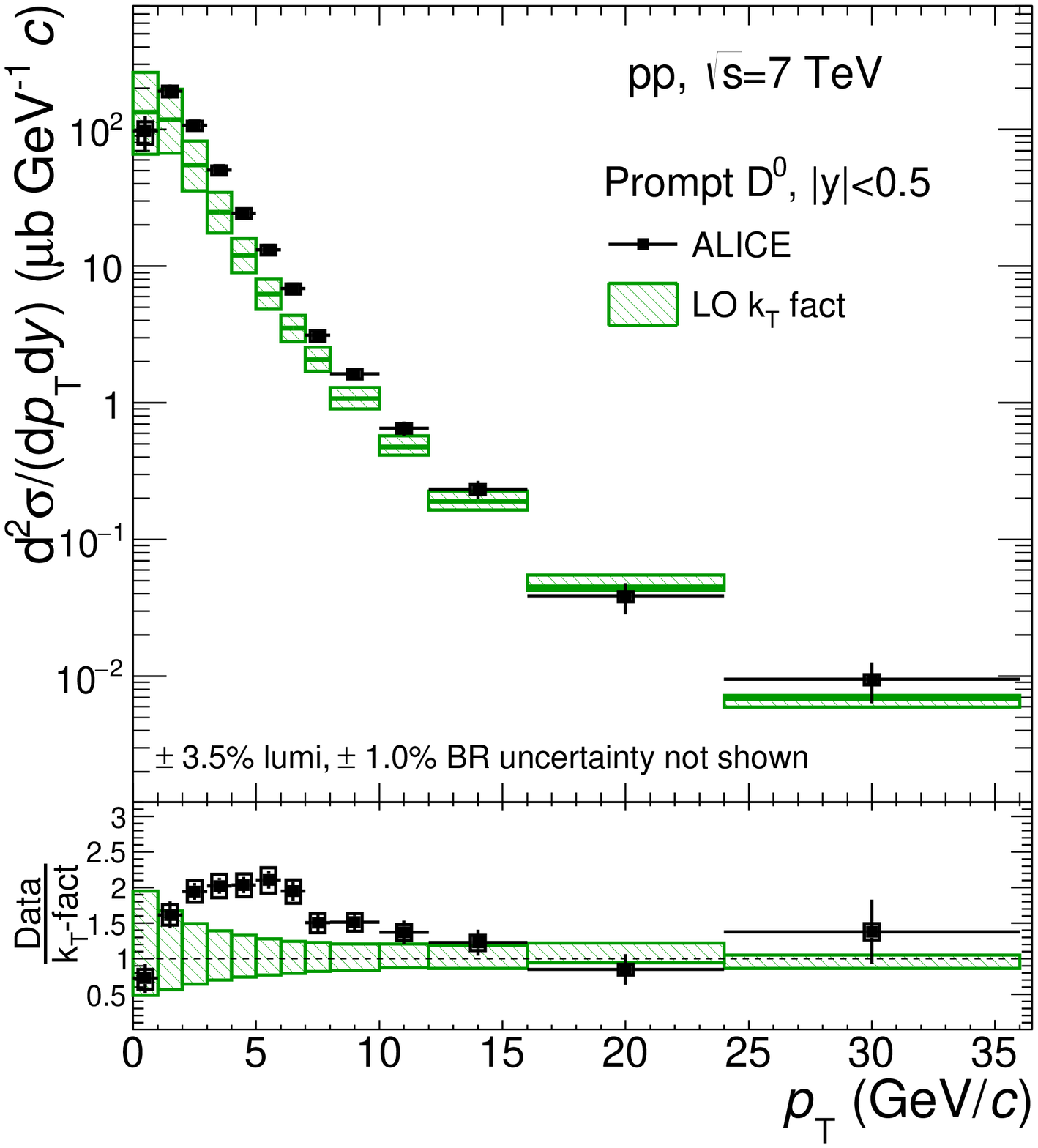}

\caption{$\pt$-differential production cross section of prompt $\Dzero$ mesons 
with $|y|<0.5$ in the interval \mbox{$0<\pt<36~\gev/c$}, in pp collisions at 
$\sqrt s =7~\tev$. 
The data point in $0<\pt<1~\gev/c$ is obtained from the analysis without decay 
vertex reconstruction described in Ref.~\cite{Adam:2016ich}.
The cross section is compared to three pQCD calculations: 
FONLL~\cite{Cacciari:2012ny} (top-left panel), GM-VFNS~\cite{Kniehl:2012ti} 
(top-right panel) and a leading order (LO) calculation based on 
$k_{\rm T}$-factorisation~\cite{Maciula:2013wg} (bottom panel).
The ratios of the data to the three calculated cross sections are shown
in the lower part of each panel. In the data-to-theory ratios the 3.5\%
normalisation uncertainty due to the luminosity determination is not included
in the systematic uncertainty on the data points.
}
\label{fig:CrossSecD0}
\end{center}
\end{figure}

\begin{figure}[!thb]
\begin{center}
\includegraphics[width=0.48\textwidth]{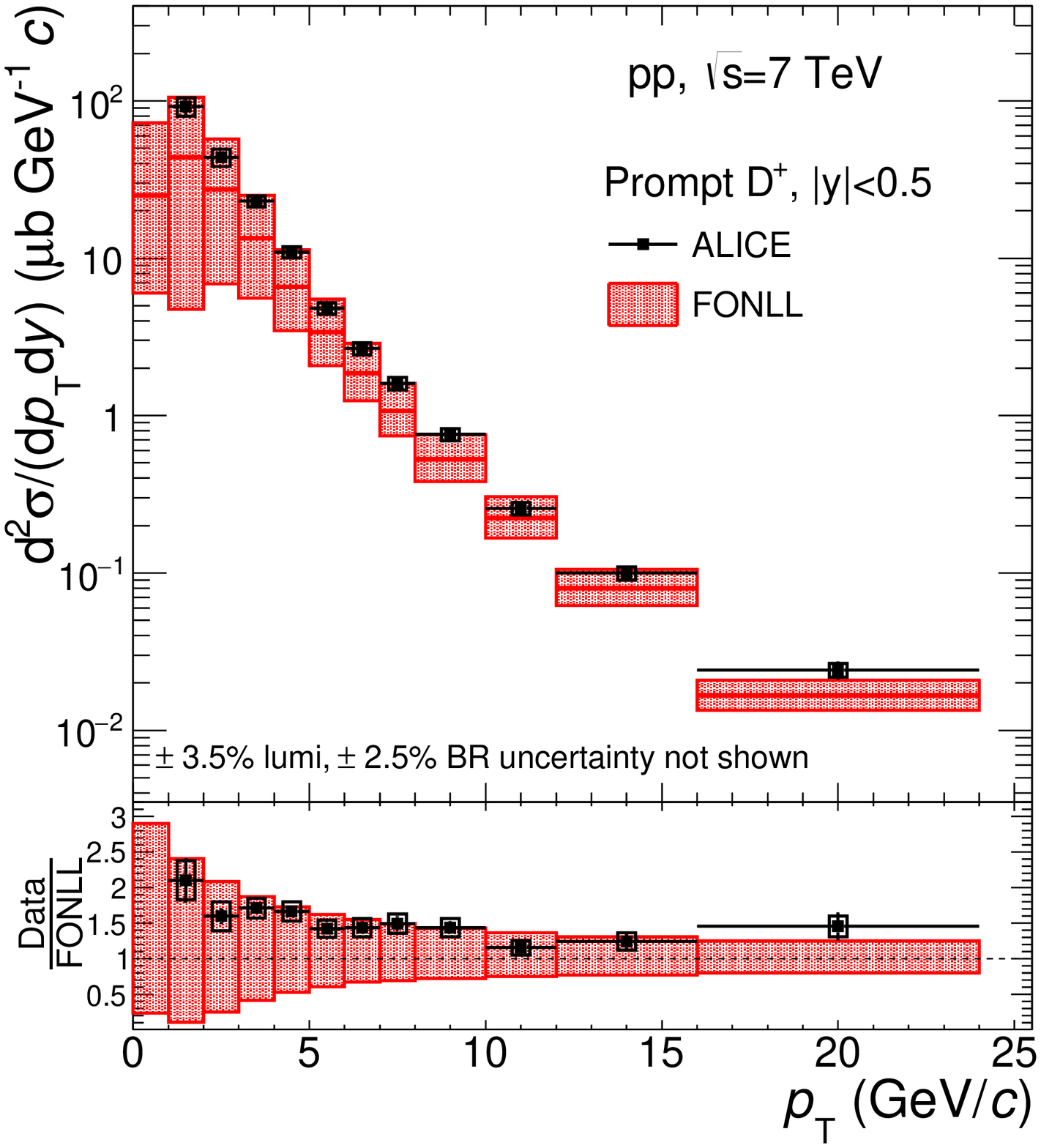}
\includegraphics[width=0.48\textwidth]{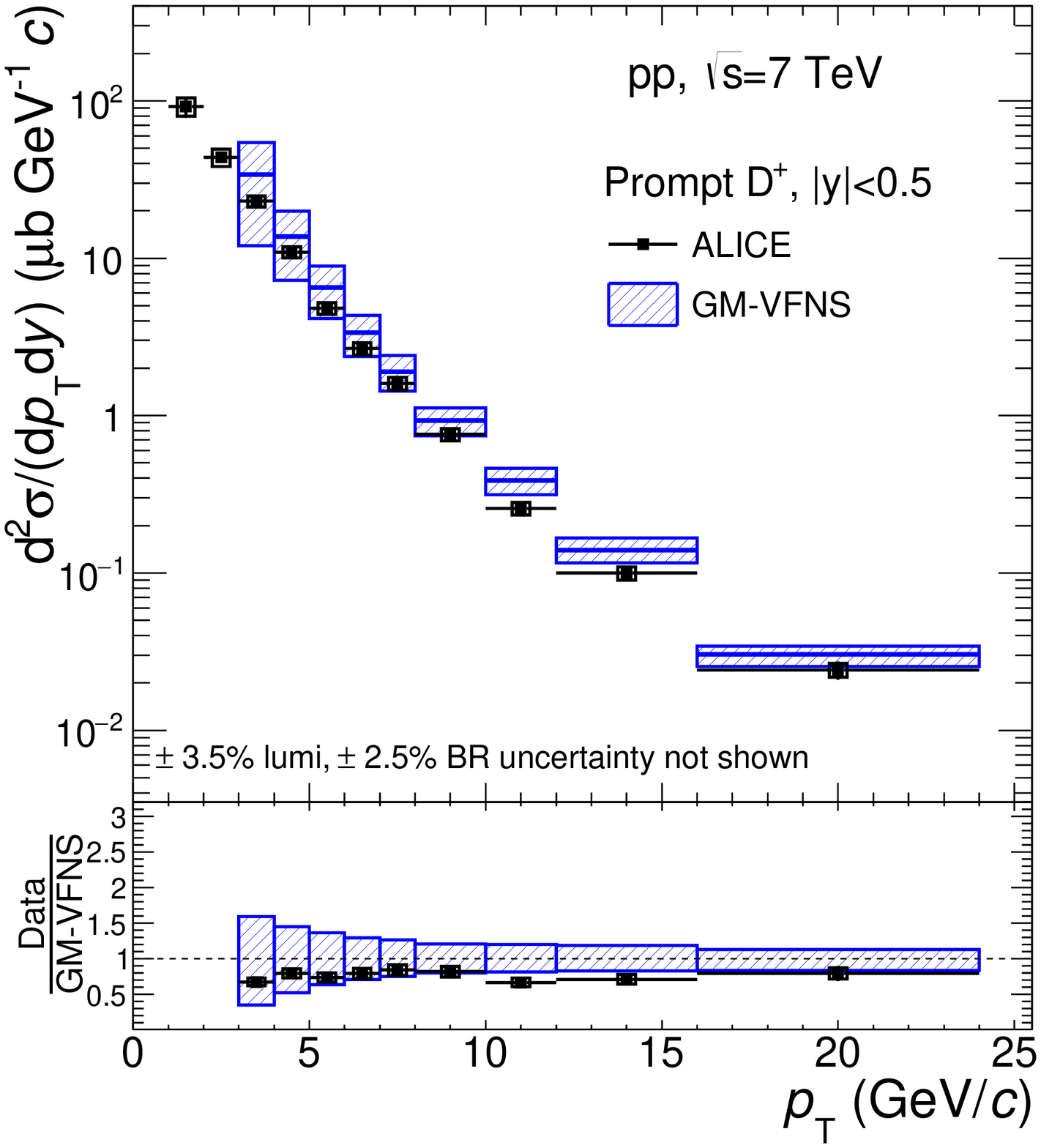}
\includegraphics[width=0.48\textwidth]{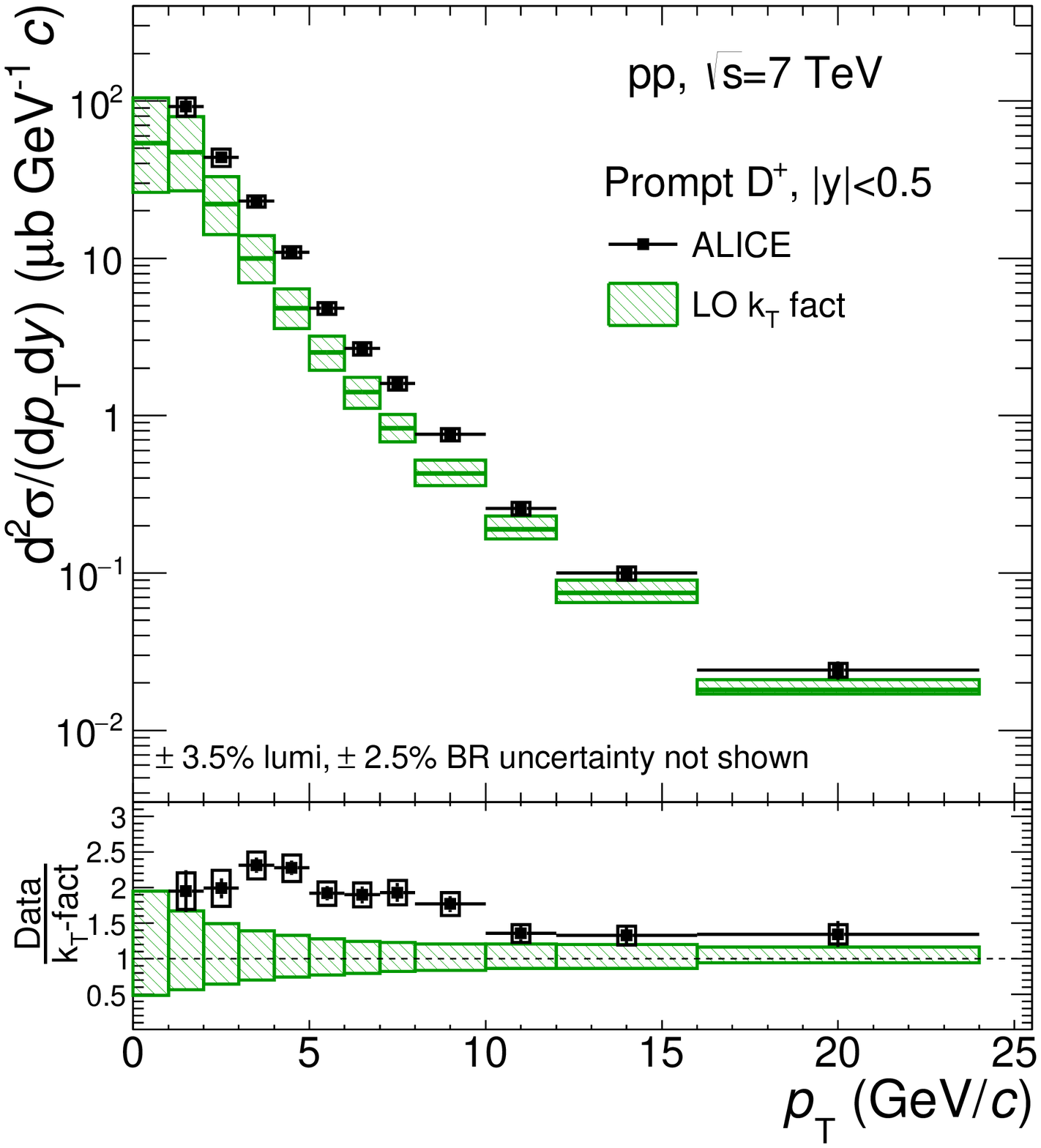}

\caption{$\pt$-differential production cross section of prompt $\Dplus$ mesons 
with $|y|<0.5$ in the interval \mbox{$1<\pt<24~\gev/c$}, in pp collisions at 
$\sqrt s =7~\tev$. 
The cross section is compared to three pQCD calculations: 
FONLL~\cite{Cacciari:2012ny} (top-left panel), GM-VFNS~\cite{Kniehl:2012ti} 
(top-right panel) and a leading order (LO) calculation based on 
$k_{\rm T}$-factorisation~\cite{Maciula:2013wg} (bottom panel).
The ratios of the data to the three calculated cross sections are shown
in the lower part of each panel. In the data-to-theory ratios the 3.5\%
normalisation uncertainty due to the luminosity determination is not included
in the systematic uncertainty on the data points.}
\label{fig:CrossSecDplus}
\end{center}
\end{figure}

\begin{figure}[!thb]
\begin{center}
\includegraphics[width=0.48\textwidth]{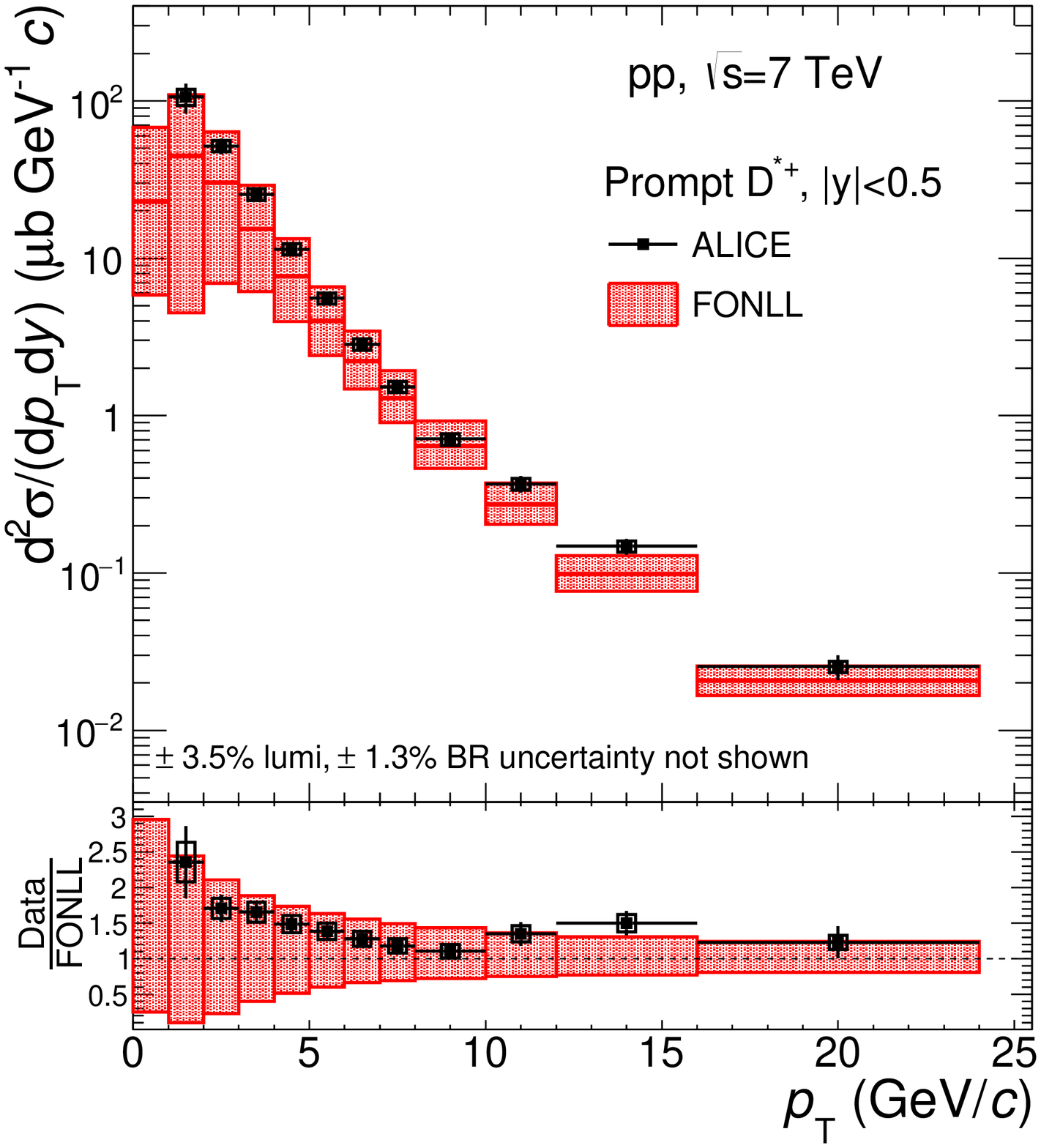}
\includegraphics[width=0.48\textwidth]{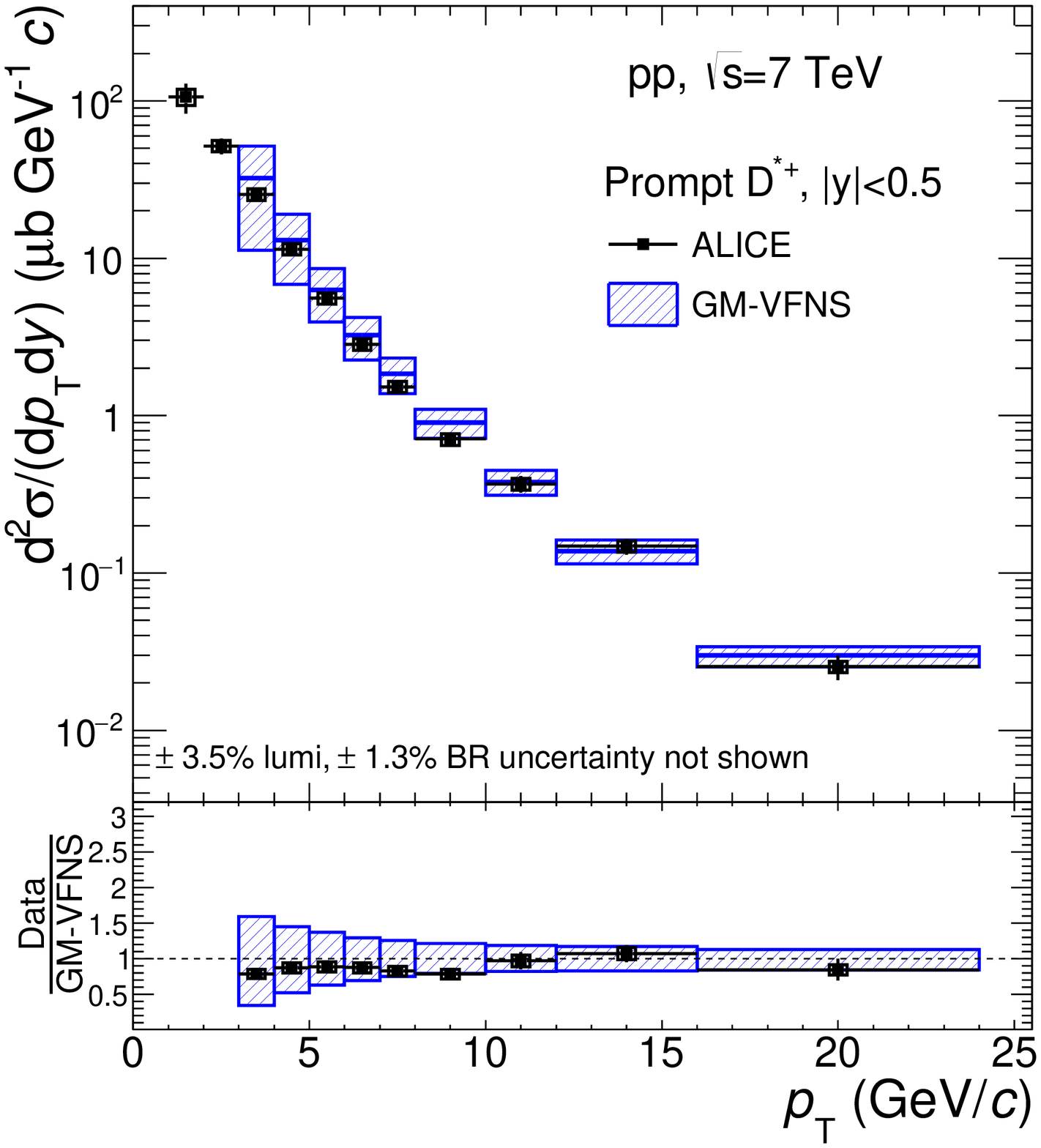}
\includegraphics[width=0.48\textwidth]{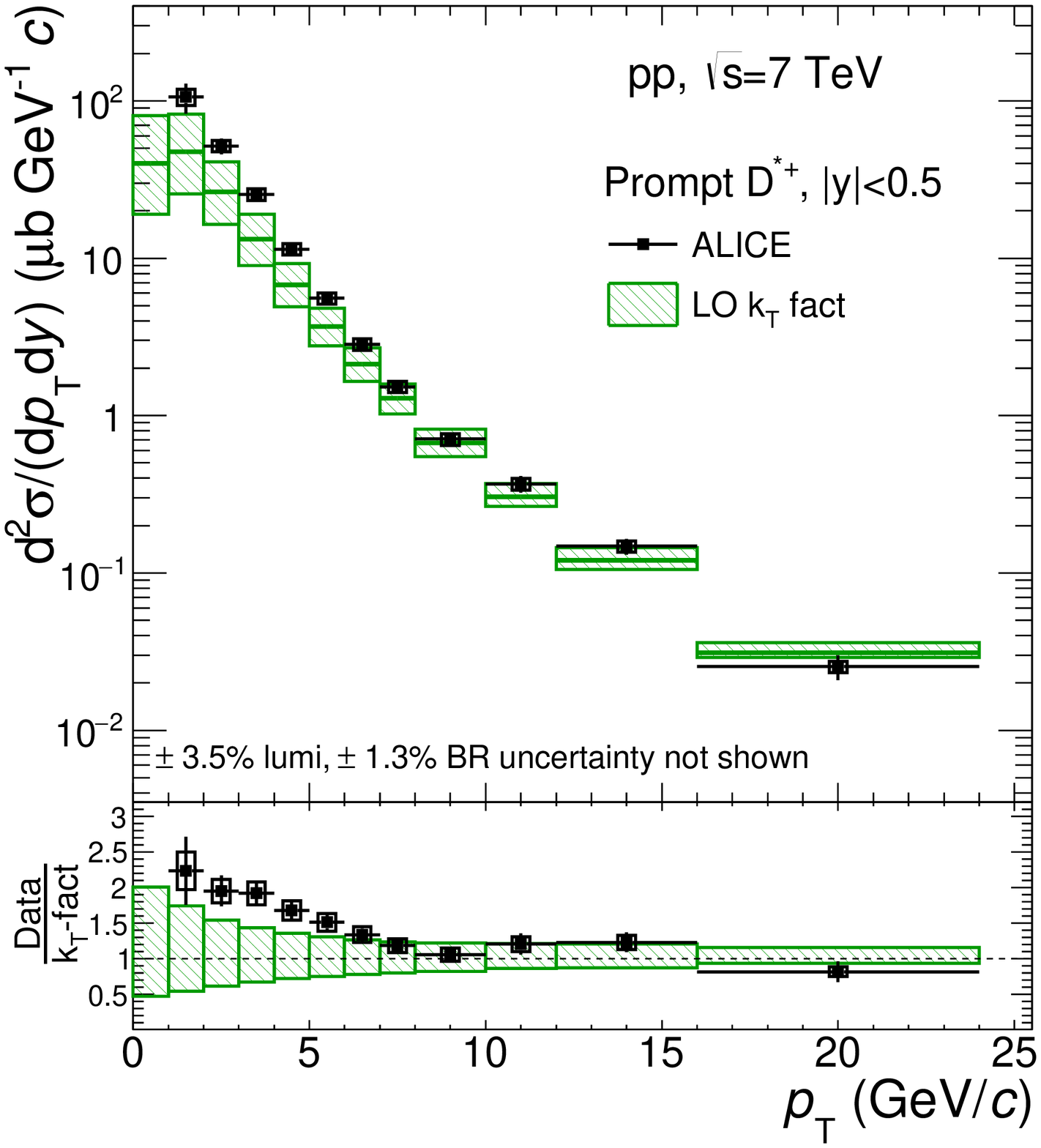}

\caption{$\pt$-differential production cross section of prompt $\Dstar$ mesons 
with $|y|<0.5$ in the interval \mbox{$1<\pt<24~\gev/c$}, in pp collisions at 
$\sqrt s =7~\tev$. 
The cross section is compared to three pQCD calculations: 
FONLL~\cite{Cacciari:2012ny} (top-left panel), GM-VFNS~\cite{Kniehl:2012ti} 
(top-right panel) and a leading order (LO) calculation based on 
$k_{\rm T}$-factorisation~\cite{Maciula:2013wg} (bottom panel).
The ratios of the data to the three calculated cross sections are shown
in the lower part of each panel. In the data-to-theory ratios the 3.5\%
normalisation uncertainty due to the luminosity determination is not included
in the systematic uncertainty on the data points.}
\label{fig:CrossSecDstar}
\end{center}
\end{figure}

\begin{figure}[!thb]
\begin{center}
\includegraphics[width=0.48\textwidth]{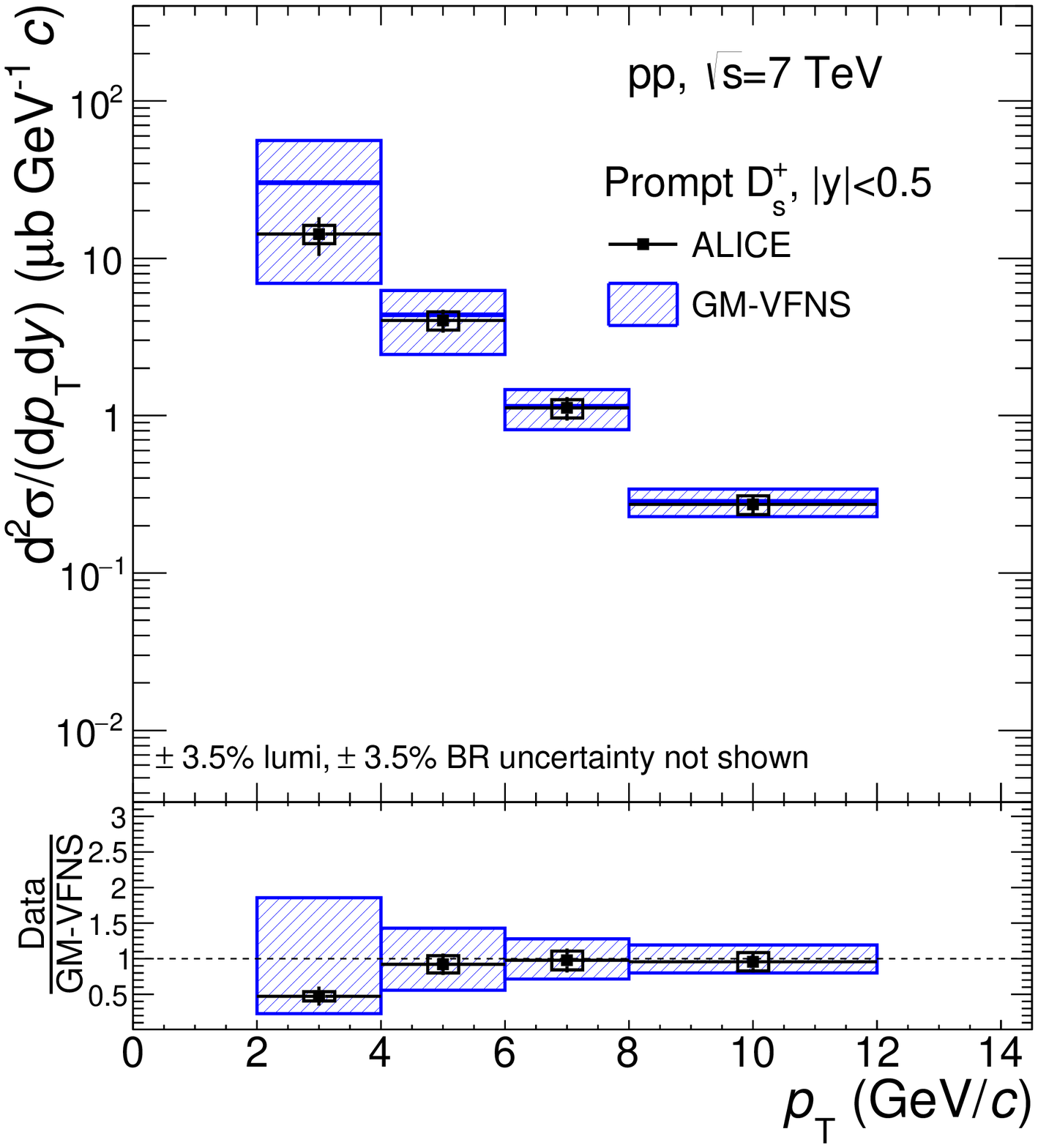}
\includegraphics[width=0.48\textwidth]{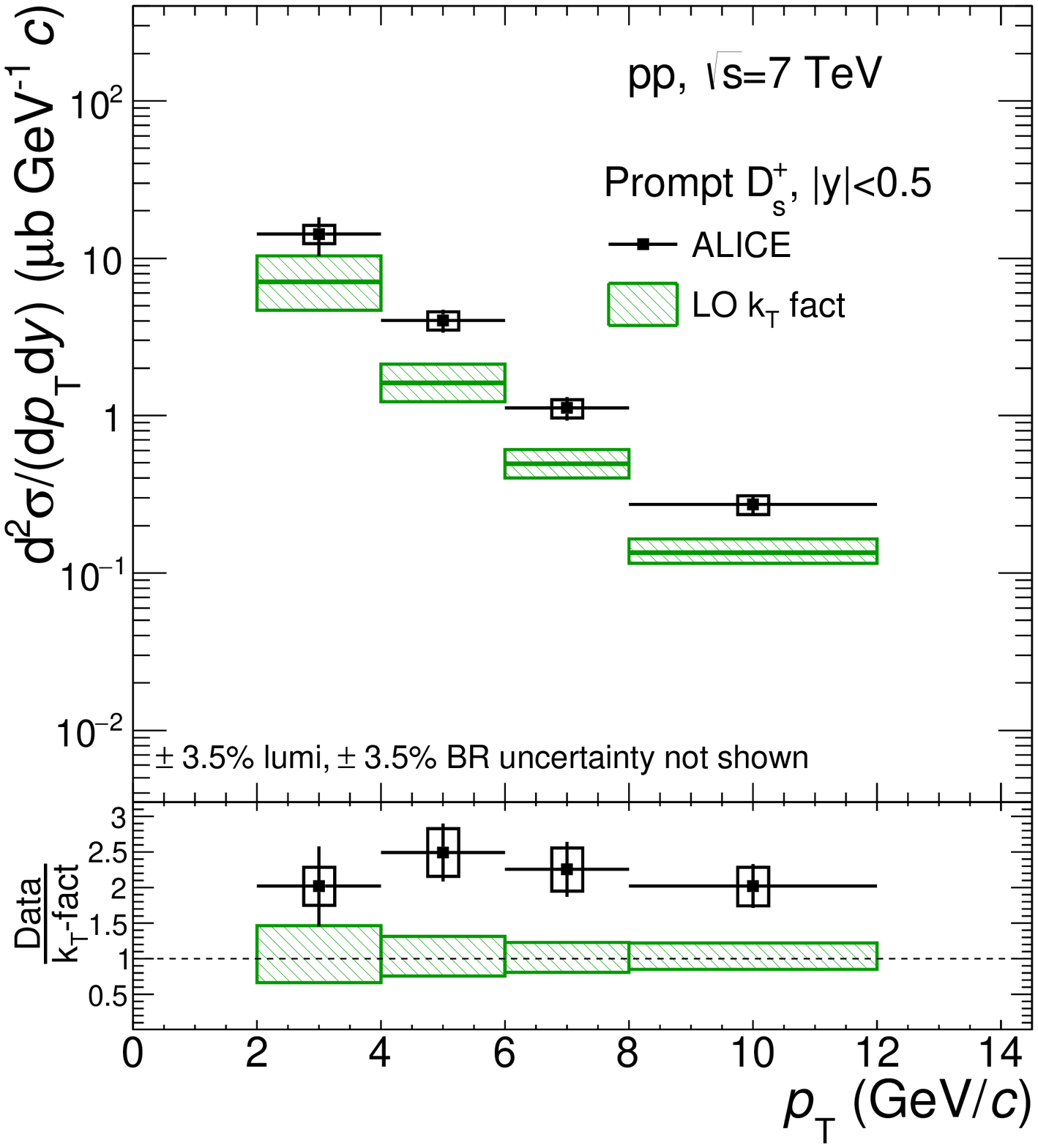}
\caption{$\pt$-differential production cross section of prompt $\Ds$ mesons 
with $|y|<0.5$ in the interval \mbox{$2<\pt<12~\gev/c$}, in pp collisions at 
$\sqrt s =7~\tev$. 
The cross section is compared to two pQCD calculations: 
GM-VFNS~\cite{Kniehl:2012ti} (left panel) and a leading order (LO) calculation 
based on $k_{\rm T}$-factorisation~\cite{Maciula:2013wg} (right panel).
The ratios of the data to the calculated cross sections are shown
in the lower part of each panel. In the data-to-theory ratios the 3.5\%
normalisation uncertainty due to the luminosity determination is not included
in the systematic uncertainty on the data points.}
\label{fig:CrossSecDs}
\end{center}
\end{figure}

In Figs.~\ref{fig:CrossSecD0}--\ref{fig:CrossSecDs}, the measured 
$\pt$-differential cross sections are compared with results from 
perturbative QCD calculations, two of which are based on collinear 
factorisation (FONLL~\cite{Cacciari:1998it,Cacciari:2012ny}
and GM-VFNS~\cite{Kniehl:2004fy,Kniehl:2005mk,Kniehl:2012ti}) and one is a 
leading order (LO) calculation based on 
$k_{\rm T}$-factorisation~\cite{Maciula:2013wg}.
The results of these calculations, performed in the same $\pt$ intervals of the 
measurement, are shown as filled boxes spanning the theoretical uncertainties
and a solid line representing the values obtained with the central values of 
the pQCD parameters.
The theoretical uncertainties are estimated in all the three frameworks
by varying the renormalisation and factorisation scales. 
In the FONLL and $k_{\rm T}$-factorisation calculations also the effect of the 
charm-quark mass uncertainty is considered.
In the FONLL and GM-VFNS calculations, the CTEQ6.6 PDFs~\cite{Pumplin:2002vw}
were used, and the uncertainty on the PDFs was included in the 
FONLL error boxes.
The LO $k_{\rm T}$-factorisation calculations were performed
with an updated set of unintegrated gluon-distribution functions computed 
from the recent MMHT2014-LO PDFs~\cite{Harland-Lang:2014zoa}.
For this reason, the comparison to the measured $\Dzero$-meson cross section 
differs from that reported in Ref.~\cite{Adam:2016ich}.
In the FONLL calculation, the fragmentation fractions $f(\rm c\to D)$,
i.e.\ the fractions of charm quarks hadronising into each D-meson species,
were taken from Ref.~\cite{Gladilin:2014tba}.
For the $\Ds$ mesons, only the comparisons to GM-VFNS and LO 
$k_{\rm T}$-factorisation predictions are shown, because a calculation of the 
$\Ds$ production cross section within the FONLL framework is not available.
The central value of the GM-VFNS predictions lies systematically 
above the data, while that of the FONLL predictions lies below the data. 
For FONLL, this feature was observed also at  other values of $\sqrt s$, from 
0.2 to 13~TeV~\cite{Adamczyk:2012af,Acosta:2003ax,Abelev:2012vra,Aaij:2013mga,Aaij:2016jht,Aaij:2015bpa}.
The LO $k_{\rm T}$-factorisation calculation describes the data within 
uncertainties for $\pt<2~\gev/c$ and $\pt>10~\gev/c$, while in the interval
$2<\pt<10~\gev/c$ the predictions underestimate the measured production
cross sections.

The average transverse momentum $\meanpt$ of prompt $\Dzero$ mesons 
was measured by fitting the cross section 
reported in Fig.~\ref{fig:CrossSecD0} with a power-law function:
\begin{equation}
f(\pt)=C \frac{\pt}{(1+(\pt/p_0)^2)^n}\,,
\label{powlaw}
\end{equation}
where $C$, $p_0$ and $n$ are the free parameters.
The prompt-$\Dzero$ $\meanpt$, defined as the mean value of the fit function, is:
\begin{equation}
\meanpt^{\rm prompt\,D^0}_{\rm pp,\,7\,TeV} = 2.19 \pm 0.06\,({\rm stat.})\, \pm 0.04\,({\rm syst.})~\gev/c\,.
\end{equation}
The systematic uncertainty on $\meanpt$ was estimated as described in 
Ref.~\cite{Adam:2016ich} taking into account separately the contributions due 
to the correlated and uncorrelated systematic uncertainties on the measured
$\pt$-differential cross section.
The uncertainty due to the fit function was estimated from the spread of the 
results obtained with different functions and using an alternative method,
which is not based on fits to the distribution, but on direct calculations 
of $\meanpt$ from the data points.

\begin{figure}[!thb]
\begin{center}
\includegraphics[width=\textwidth]{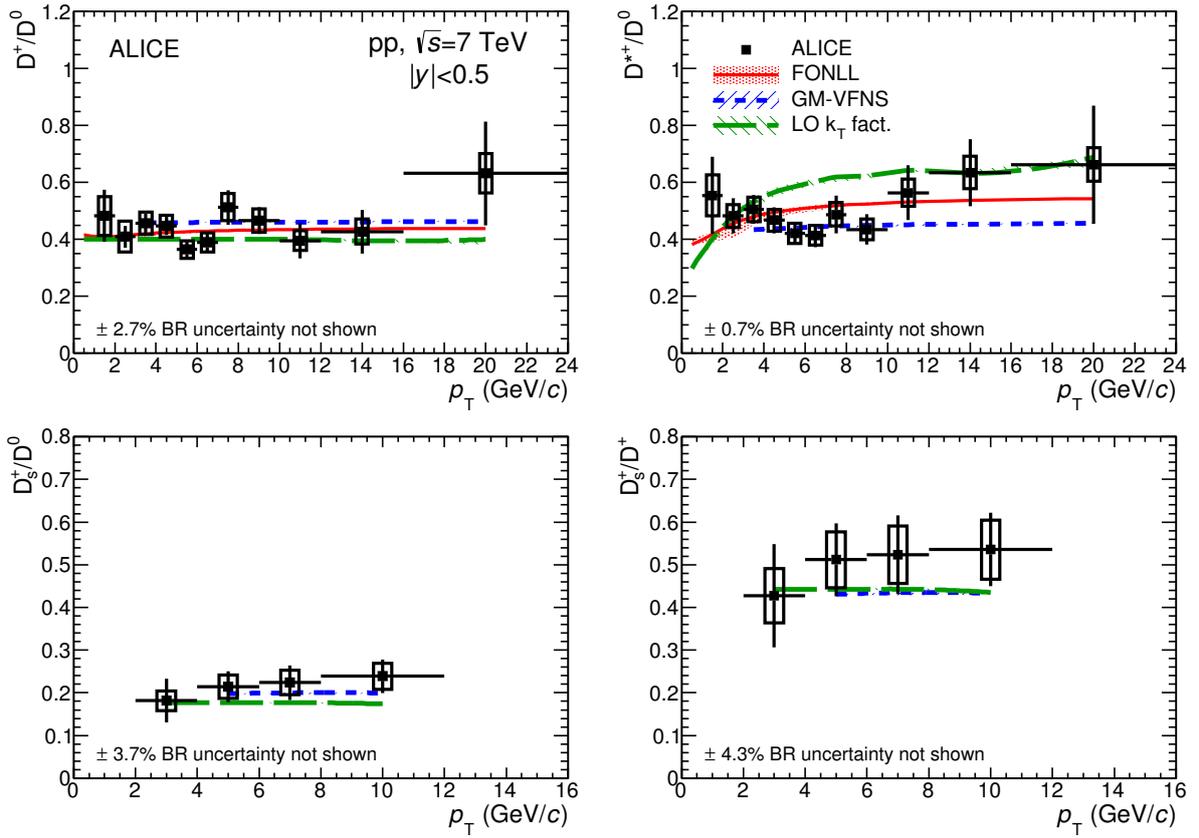}

\caption{Ratios of D-meson production cross sections as a function of $\pt$.
Predictions from FONLL, GM-VFNS and LO $\kt$-factorisation calculations
are also shown. For the pQCD calculations the line shows the ratio of
the central values of the theoretical cross sections, while the shaded area is 
defined by the ratios computed from the upper and lower limits of the 
theoretical uncertainty band.}
\label{fig:DratiosVsPt}
\end{center}
\end{figure}

The ratios of the $\pt$-differential cross sections of $\Dzero$, $\Dplus$, 
$\Dstar$ and $\Ds$ mesons are reported in Fig.~\ref{fig:DratiosVsPt}.
In the evaluation of the systematic uncertainties on these ratios, 
the sources of correlated and uncorrelated systematic effects were treated 
separately. 
In particular, the contributions of the yield extraction and cut efficiency 
were considered as uncorrelated, while those of the feed-down from 
beauty-hadron decays and the tracking efficiency were treated as fully 
correlated among the different D-meson species.
The measured D-meson ratios do not show a significant $\pt$ dependence within 
the experimental uncertainties, thus suggesting a small difference 
between the fragmentation functions of charm quarks to pseudoscalar 
($\Dzero$, $\Dplus$ and $\Ds$) and vector ($\Dstar$) mesons and to strange and 
non-strange mesons.
The data are compared to the ratios of the D-meson cross sections from FONLL 
(only for $\Dzero$, $\Dplus$ and $\Dstar$ mesons), GM-VFNS and LO 
$\kt$-factorisation pQCD calculations.
The ratios of the theoretical predictions were computed assuming their 
uncertainties to be fully correlated among the D-meson species, which 
results in an almost complete cancellation of the uncertainties in the ratio. 
Note that in all these pQCD calculations, the relative abundances of the 
different D-meson species are not predicted by the theory, but 
the fragmentation fractions, $f({\rm c\rightarrow D})$, are taken from 
the experimental measurements~\cite{Kneesch:2007ey,Cacciari:2012ny,Cacciari:2003zu,Maciula:2013wg,Barate:1999bg,Gladilin:2014tba}.
In the FONLL and GM-VFNS frameworks, the $\pt$ dependence 
of the ratios of the D-meson production cross sections arises from the 
different fragmentation functions used to model the transfer of energy from 
the charm quark to a specific D-meson 
species~\cite{Cacciari:2003zu,Kneesch:2007ey,Kniehl:2006mw}, 
and from the different contribution from decays of higher excited states.
The parton fragmentation models used in the calculations provide
an adequate description of the measured data.
In the LO $\kt$-factorisation calculations, the same fragmentation function 
(Peterson~\cite{Peterson:1982ak}) is used for $\Dzero$, $\Dplus$ and $\Ds$ 
mesons, resulting in the same shape of the $\pt$ distributions of these three 
meson species, while 
the fragmentation functions for vector mesons from Ref.~\cite{Braaten:1994bz} 
are used for $\Dstar$ mesons~\cite{Maciula:2013wg}.

\begin{figure}[!tb]
\begin{center}
\includegraphics[width=\textwidth]{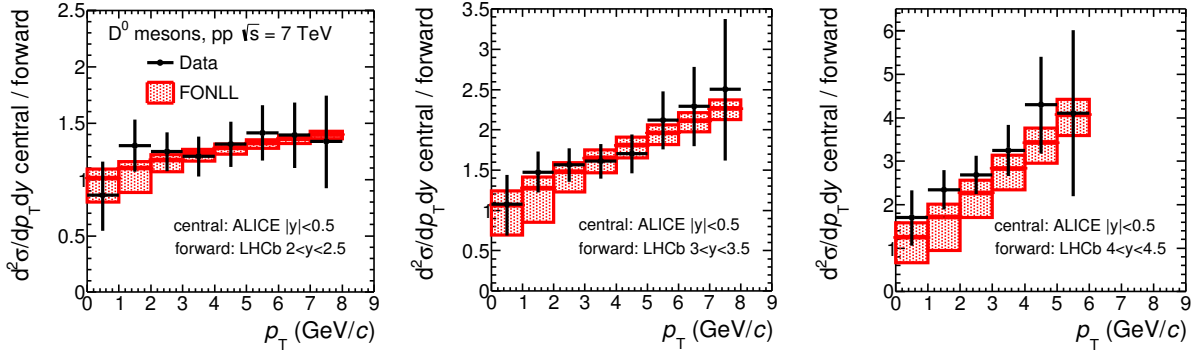}
\caption{Ratios of $\Dzero$-meson production cross section per unit of rapidity at mid-rapidity ($|y|<0.5$) to that measured by the LHCb 
Collaboration~\cite{Aaij:2013mga} in three 
rapidity ranges, $2<y<2.5$ (left panel), 
$3<y<3.5$ (middle panel), $4<y<4.5$ (right panel), as a function 
of $\pt$. The LHCb measurement were multiplied by 2 to refer them to one unit of rapidity. 
The error bars represent the total (statistical and systematic) uncertainty 
on the measurement.
Predictions from FONLL calculations are compared to the data points.}
\label{fig:CentForwRatios}
\end{center}
\end{figure}

The ratios of $\Dzero$-meson production cross sections in different rapidity 
intervals, which are expected to be sensitive to the gluon PDF at small 
values of Bjorken-$x$~\cite{Cacciari:2015fta}, were computed from our
measurement in the central rapidity region ($|y|<0.5$) and the results reported
by the LHCb collaboration for pp collisions at $\sqrts=7~\TeV$ 
in different $y$ intervals at forward rapidity~\cite{Aaij:2013mga}.
The results are reported in Fig.~\ref{fig:CentForwRatios}, where the 
central-to-forward ratios are shown as a function of $\pt$ for three 
different $y$ intervals at forward rapidity: $2<y<2.5$ (left panel), 
$3<y<3.5$ (middle panel), $4<y<4.5$ (right panel).
The error bars represent the total uncertainty obtained from the propagation 
of the statistical and systematic uncertainties on the $\pt$-differential 
cross sections.
The measured ratios are compared to FONLL calculations, shown as boxes in
Fig.~\ref{fig:CentForwRatios}.
The central-to-forward ratios computed using the central values of the 
factorisation and renormalisation scales are found to describe the data within 
their uncertainties.
The upper edge of the FONLL uncertainty band, which is also in agreement
with the measured values of the central-to-forward ratios, is determined by 
the calculations with factorisation scale $\mu_{\rm F}=2\,m_{\rm T}$,
where $m_{\rm T}=\sqrt{\pt^2+m_{\rm c}^2}$ and $m_{\rm c}=1.5~\gev/c^2$.
The low edge of the FONLL uncertainty band is determined by the 
calculations with $\mu_{\rm F}=0.5\,m_{\rm T}$, which provide a worse description 
of the measured central-to-forward ratios at low $\pt$ for the most forward 
rapidity interval.
Note that in this forward rapidity interval, the FONLL calculation with 
$\mu_{\rm F}=0.5\,m_{\rm T}$ uses the PDFs for Bjorken-$x$ values reaching 
down to about $10^{-5}$, a region that is not constrained by experimental data, 
and below $Q^2 = (1.3~\GeV/c)^2$, where the CTEQ6.6 PDFs~\cite{Pumplin:2002vw}
are kept constant to their values at $(1.3~\GeV/c)^2$.

The visible cross sections of prompt D mesons, obtained by integrating
the $\pt$-differential cross sections in the measured $\pt$ range, are reported
in Table~\ref{tab:ptintegcs}.
In addition, for $\Dzero$ mesons the cross sections integrated over 
the $\pt$ intervals of the $\Dplus$, $\Dstar$ and $\Ds$ measurements are 
shown.
The systematic uncertainty was defined by propagating the yield extraction uncertainties as uncorrelated among $\pt$ intervals and all the other uncertainties 
as correlated.
These values were used to compute the ratios of the $\pt$-integrated D-meson 
production cross sections, which are reported in Table~\ref{tab:ptintegrat}.
The systematic uncertainties on the ratios were computed taking into
account the sources correlated and uncorrelated among the different D-meson
species as described above.
The measured ratios are compatible within uncertainties with the results at
$\sqrts=2.76~\TeV$~\cite{Abelev:2012vra} and with the measurements of
the LHCb collaboration at forward rapidity ($2.0<y<4.5$) at three 
different collision energies $\sqrts=$~5, 7 and 
13~TeV~\cite{Aaij:2013mga,Aaij:2016jht,Aaij:2015bpa}.
The measured $\pt$-integrated production ratios are also compatible with the
charm-quark fragmentation fractions $f(\rm c\to D)$ measured in ${\rm e^+e^-}$ 
collisions from the compilation in~\cite{Gladilin:2014tba}.
These results indicate that the fragmentation fractions of charm quarks into
different D-meson species do not vary substantially with rapidity, collision
energy and colliding system.

\begin{table}
\centering
\begin{tabular}{c|c|l} 
 & Kinematic range & Visible cross section ($\mub$) \\
\hline
\rule{0pt}{12pt} 
$\Dzero$  & $0<\pt<36~\GeV/c$ & $500 \pm 36 ({\rm stat}) \pm 39 ({\rm syst}) \pm 18 ({\rm lumi}) \pm 5 ({\rm BR})$ \\[1ex]
          & $1<\pt<24~\GeV/c$ & $402 \pm 24 ({\rm stat}) \pm 28 ({\rm syst}) \pm 14 ({\rm lumi}) \pm 4 ({\rm BR})$ \\[1ex]
          & $2<\pt<12~\GeV/c$ & $210 \pm \phantom{0}7 ({\rm stat}) \pm 14 ({\rm syst}) \pm \phantom{0}7 ({\rm lumi}) \pm 2 ({\rm BR})$\\[1ex]
\hline
\rule{0pt}{12pt} 
$\Dplus$  & $1<\pt<24~\GeV/c$ & $182 \pm 14 ({\rm stat}) \pm 20 ({\rm syst}) \pm \phantom{0}6 ({\rm lumi}) \pm 5 ({\rm BR})$\\[1ex]
          & $2<\pt<12~\GeV/c$ & $\phantom{0}89 \pm \phantom{0}3 ({\rm stat}) \pm \phantom{0}9 ({\rm syst}) \pm \phantom{0}3 ({\rm lumi}) \pm 2 ({\rm BR})$\\[1ex]
\hline
\rule{0pt}{12pt} 
$\Dstar$  & $1<\pt<24~\GeV/c$ & $207 \pm 24 ({\rm stat}) \pm 20 ({\rm syst}) \pm \phantom{0}7 ({\rm lumi}) \pm 3 ({\rm BR})$\\[1ex]
          & $2<\pt<12~\GeV/c$ & $101 \pm \phantom{0}6 ({\rm stat}) \pm \phantom{0}8 ({\rm syst}) \pm \phantom{0}4 ({\rm lumi}) \pm 1 ({\rm BR})$\\[1ex]
\hline
\rule{0pt}{12pt} 
$\Ds$     & $2<\pt<12~\GeV/c$ & $\phantom{0}40 \pm \phantom{0}8 ({\rm stat}) \pm \phantom{0}5 ({\rm syst}) \pm \phantom{0}1 ({\rm lumi}) \pm 1 ({\rm BR})$\\[1ex]
\hline
\end{tabular}
\caption{Visible production cross sections of prompt D mesons in $|y| < 0.5$ in pp collisions at $\sqrts = 7~\TeV$.}
\label{tab:ptintegcs}
\end{table}

\begin{table}
\centering
\begin{tabular}{c|c|l} 
 & Kinematic range & Production cross section ratio \\
\hline
\rule{0pt}{12pt} 
$\sigma(\Dplus) / \sigma(\Dzero)$ & $1<\pt<24~\GeV/c$ & $0.45 \pm 0.04 ({\rm stat}) \pm 0.05 ({\rm syst}) \pm 0.01 ({\rm BR})$\\[1ex]
$\sigma(\Dstar) / \sigma(\Dzero)$ & $1<\pt<24~\GeV/c$ & $0.52 \pm 0.07 ({\rm stat}) \pm 0.05 ({\rm syst}) \pm 0.01 ({\rm BR})$\\[1ex]
$\sigma(\Ds) / \sigma(\Dzero)$    & $2<\pt<12~\GeV/c$ & $0.19 \pm 0.04 ({\rm stat}) \pm 0.02 ({\rm syst}) \pm 0.01 ({\rm BR})$\\[1ex]
$\sigma(\Ds) / \sigma(\Dplus)$    & $2<\pt<12~\GeV/c$ & $0.45 \pm 0.09 ({\rm stat}) \pm 0.06 ({\rm syst}) \pm 0.02 ({\rm BR})$\\[1ex]
\hline
\end{tabular}
\caption{Ratios of the measured $\pt$-integrated cross sections of prompt D mesons in $|y| < 0.5$ in pp collisions at $\sqrts = 7~\TeV$.}
\label{tab:ptintegrat}
\end{table}

The production cross sections per unit of rapidity, ${\rm d}\sigma/{\rm d}y$, 
at mid-rapidity were computed for each meson species by extrapolating the 
visible cross section to the full $\pt$ range.
The extrapolation factor for a given D-meson species was defined as
the ratio between the total production cross section in $|y|<0.5$ and that 
in the experimentally covered phase space, both of them calculated with the
FONLL central parameters. 
The systematic uncertainty on the extrapolation factor was estimated by
considering the contributions due to i) the uncertainties on the 
CTEQ6.6 PDFs~\cite{Pumplin:2002vw} and ii) the variation of the charm-quark 
mass and the renormalisation and factorisation scales in the FONLL 
calculation, as proposed in~\cite{Cacciari:2012ny}.
For $\Dzero$ mesons, which are measured down to $\pt=0$, the extrapolation
factor accounts only for the very small contribution of D-mesons with
$\pt>36~\GeV/c$ and it has therefore a value very close to unity with 
negligible uncertainty.
In the case of $\Ds$ mesons, for which a FONLL prediction is not available,
the central value of the extrapolation factor was computed 
from the prediction based on the $\pt$-differential cross section of charm 
quarks from FONLL, the fractions $f({\rm c\rightarrow \Ds})$ and 
$f({\rm c\rightarrow D_{\rm s}^{*+}})$ from ALEPH~\cite{Barate:1999bg}, and the 
fragmentation functions from~\cite{Braaten:1994bz}, which have one parameter, 
$r$, that was set to 0.1 as done in FONLL~\cite{Cacciari:2003zu}.
The ${\rm D}_{\rm s}^{*+}$ mesons produced in the c quark fragmentation were 
made to decay with PYTHIA and the resulting $\Ds$ mesons were summed to the 
primary ones to obtain the prompt yield.
An additional contribution to the systematic uncertainty  was assigned
for $\Ds$ mesons based on the envelope of the results obtained using the 
FONLL $\pt$-differential cross sections of $\Dzero$, 
$\Dplus$ and $\Dstar$ mesons to compute the $\Ds$ extrapolation factor.
The resulting values for the extrapolation factors and for the prompt D-meson 
production cross sections per unit of rapidity ${\rm d}\sigma/{\rm d}y$ are 
reported in Table~\ref{tab:dsdy}.

\begin{table}
\centering
\begin{tabular}{c|c|l} 
 & Extr. factor to $\pt>0$ & ${\rm d}\sigma/{\rm d}y \mid_{|y|<0.5}$ ($\mub$) \\
\hline
\rule{0pt}{12pt} 
$\Dzero$  & $1.0002 ^{+0.0004}_{-0.0002}$ & $500 \pm 36 ({\rm stat}) \pm 39 ({\rm syst}) \pm 18 ({\rm lumi}) \pm 5 ({\rm BR})$\\[1ex]
\hline
\rule{0pt}{12pt} 
$\Dplus$  & $1.25 ^{+0.29}_{-0.09}$ & $227 \pm 18 ({\rm stat}) \pm 25 ({\rm syst}) \pm \phantom{0}8 ({\rm lumi}) \pm 6 ({\rm BR})^{+52}_{-16} ({\rm extrap})$ \\[1ex]
\hline
\rule{0pt}{12pt} 
$\Dstar$  & $1.21 ^{+0.28}_{-0.08}$ & $251 \pm 29 ({\rm stat}) \pm 24 ({\rm syst}) \pm \phantom{0}9 ({\rm lumi}) \pm 3 ({\rm BR})^{+58}_{-16} ({\rm extrap})$\\[1ex]
\hline
\rule{0pt}{12pt} 
$\Ds$     & $2.23^{+0.71}_{-0.65}$ &  $\phantom{0}89 \pm 18 ({\rm stat}) \pm 11 ({\rm syst}) \pm \phantom{0}3 ({\rm lumi}) \pm 3 ({\rm BR})^{+28}_{-26} ({\rm extrap})$\\[1ex]
\hline
\end{tabular}
\caption{Production cross sections of prompt D mesons in $|y| < 0.5$ and full $\pt$ range in pp collisions at $\sqrts = 7~\TeV$.}
\label{tab:dsdy}
\end{table}

The ${\rm c\overline{c}}$ production cross section per unit of rapidity at mid-rapidity ($|y|<0.5$)
was calculated by dividing the prompt $\Dzero$-meson cross section
by the fraction of charm quarks hadronising into $\Dzero$ mesons 
$f(c\rightarrow\Dzero)$ and correcting for the different shapes of the 
distributions of $y_{\Dzero}$ and $y_{\rm c\overline{c}}$ (${\rm c\overline{c}}$ 
pair rapidity).
The correction factor and its uncertainty were extracted from FONLL and
MNR NLO pQCD~\cite{Mangano:1991jk} calculations together with 
PYTHIA\,6~\cite{Sjostrand:2006za} and POWHEG~\cite{Frixione:2007nw} 
simulations, as described in detail in Ref.~\cite{Adam:2016ich}.
For the fragmentation fraction, the value 
$f(c\rightarrow\Dzero)=0.542 \pm 0.024$ derived in Ref.~\cite{Gladilin:2014tba} 
by averaging the measurements from e$^+$e$^-$ collisions at LEP was used. 
As pointed out in Refs.~\cite{Abramowicz:2013eja,Lisovyi:2015uqa}, 
measurements in e$^+$e$^-$, ep and pp collisions agree within uncertainties, 
supporting the hypothesis that fragmentation is independent of the specific 
production process\footnote{In Ref.~\cite{Lisovyi:2015uqa}, 
an average of the charm fragmentation fractions over the measurements from 
all collision systems is calculated, imposing the constraint that the sum of 
all weakly-decaying charmed hadrons is unity, which results in 
$f(c\rightarrow\Dzero)=0.6086 \pm 0.0076$ 
(about 11\% larger that the value from~\cite{Gladilin:2014tba}).}.
The resulting  ${\rm c\overline{c}}$ cross section per unit of rapidity at 
mid-rapidity is:
\begin{equation}
\left. {\rm d}\sigma^{\rm c\overline{c}}_{\rm pp,\,7\,TeV}/{\rm d}y \, \right|_{|y|<0.5}
\, = \,  954 \pm 69 \,({\rm stat}) \pm 74\,({\rm syst}) \pm 33\,({\rm lumi}) \pm 42\,({\rm FF}) \pm 31\,({\rm rap.shape})~\mu{\rm b}\,.
\end{equation}
We verified that the precision of the ${\rm c\overline{c}}$ production 
cross-section determination does not improve if the results calculated from
$\Dplus$, $\Dstar$ and $\Ds$ mesons, which have significantly larger extrapolation 
uncertainties as compared to the $\Dzero$ one, are included via a weighted 
average procedure, as done in Ref.~\cite{Abelev:2012vra}.
The total production cross section of prompt $\Dzero$ mesons 
(average of particles and antiparticles) was calculated
by extrapolating to full phase space the cross section measured at mid-rapidity.
The extrapolation factor was defined as the ratio of the $\Dzero$ 
production cross sections in full rapidity and in $|y|<0.5$ calculated with 
the FONLL central parameters: $8.56 ^{+2.51}_{-0.42}$.
The systematic uncertainty on the extrapolation factor was estimated
with the same procedure described above for the $\pt$ extrapolation.
The resulting cross section is:
\begin{equation}
{\rm \sigma^{\rm prompt\,D^0}_{\rm pp,\,7\,TeV}= 4.28 \pm 0.31\,({\rm stat}) \, \pm 0.33\,({\rm syst})\,^{+1.26}_{-0.24}\,({\rm extr.}) \pm 0.15\,({\rm lumi}) \pm 0.04\,({\rm BR})~{\rm mb}\,.}
\end{equation}
The total charm production cross section was calculated by dividing the 
total prompt $\Dzero$-meson production cross section by 
the fragmentation fraction reported above.
The resulting ${\rm c\overline{c}}$ production cross section in pp collisions at $\sqrt{s}=7~\TeV$ is:
\begin{equation}
{\rm \sigma^{\rm c\overline{c}}_{\rm pp,\,7\,TeV}({\rm ALICE})= 7.89 \pm 0.57 \,({\rm stat.}) \,\pm 0.61\,({\rm syst.})  \,^{+2.32}_{-0.45} ({\rm extr.}) \pm 0.28\,({\rm lumi.})\, \pm 0.35\,({\rm FF})~{\rm mb}\,,}
\end{equation}
which is consistent with the value of Ref.~\cite{Adam:2016ich} but
has smaller statistical and systematic uncertainties.
It is also compatible within uncertainties with the total charm production 
cross section reported by the ATLAS collaboration~\cite{Aad:2015zix}, which 
is calculated from $\Dplus$ and $\Dstar$ measurements in $|\eta|<2.1$ and 
$\pt>3.5~\GeV/c$ and has larger uncertainties on the extrapolation to full 
kinematic phase space as compared to our result.

A more precise determination of the ${\rm c\overline{c}}$ production cross 
section can be obtained by summing our measurement of the prompt $\Dzero$-meson 
cross section in $|y|<0.5$ and the LHCb result in $2<y<4.5$ for 
$0<\pt<8~\GeV/c$~\cite{Aaij:2013mga}, and extrapolating to full rapidity and 
$\pt$ via the ratio of FONLL calculations of the cross sections in full phase 
space and in the measured $y$ and $\pt$ intervals exploiting the symmetry around $y=0$.
The result for the ${\rm c\overline{c}}$ production cross section is:
\begin{equation}
{\rm \sigma^{\rm c\overline{c}}_{\rm pp,\,7\,TeV}({\rm ALICE,LHCb})= 7.44 \pm 0.14 \,({\rm stat.}) \,\pm 0.46\,({\rm syst.})  \,^{+0.13}_{-0.07} ({\rm extr.}) \, \pm 0.33\,({\rm FF})~{\rm mb}\,,}
\end{equation}
where the +0.11~mb extrapolation uncertainty is determined by FONLL calculations 
with factorisation scale $\mu_{\rm F}=0.5 \,m_{\rm T}$, which do not describe the 
measured central-to-forward ratios of Fig.~\ref{fig:CentForwRatios}. 
If this $\mu_{\rm F}$ value is not considered, the extrapolation uncertainty 
is reduced to $\pm0.05$~mb.

\section{Summary}
\label{sec:summary}

We have presented a new measurement of the inclusive $\pt$-differential
production cross sections of prompt $\Dzero$, $\Dplus$, $\Dstar$ and $\Ds$ 
mesons at mid-rapidity ($|y|<0.5$) in pp collisions at a centre-of-mass energy of
$\sqrt{s}=7~\tev$.
The measurements cover the transverse-momentum interval $0<\pt<36~\GeV/c$
for $\Dzero$ mesons, $1<\pt<24~\GeV/c$ for $\Dplus$ and $\Dstar$ mesons, 
and $2<\pt<12~\GeV/c$ for $\Ds$ mesons.
As compared to previously published results based on the same data 
sample~\cite{ALICE:2011aa,Abelev:2012tca}, the present results have
an extended $\pt$ coverage and total uncertainties reduced by a factor of 
about 1.5--4 depending on the D-meson species and $\pt$.
The measurements cover complementary ranges in $\pt$ and $y$ with respect to those of the ATLAS
($3.5<\pt<100~\gev/c$, $|\eta|<2.1$~\cite{Aad:2015zix}) and LHCb ($0<\pt<8~\gev/c$, $2<y<4.5$~\cite{Aaij:2013mga}) Collaborations
at the same centre-of-mass energy.
The $\pt$-differential cross sections are described within uncertainties in the full $\pt$ range 
by the FONLL and GM-VFNS perturbative QCD calculations, which are based on collinear factorisation, 
while a leading-order calculation based on $k_{\rm T}$ factorisation underestimates the measured cross sections for $2<\pt<10~\gev/c$.
The $\pt$-differential ratios of our measurement at mid-rapidity and
LHCb measurements at forward rapidity~\cite{Aaij:2013mga} are described
by FONLL calculations. 
These central-to-forward ratios, once complemented with similar measurements 
at different centre-of-mass energies, could provide sensitivity to the gluon 
PDF at small values of Bjorken-$x$~\cite{Cacciari:2015fta}.
The ratios of the cross sections of the four D-meson species were found to
be compatible with the LHCb measurements at forward rapidity and different
collision energies as well as with results from ${\rm e^+e^-}$ collisions,
indicating that the fragmentation fractions of charm quarks into
different D-meson species do not vary substantially with rapidity, collision
energy and colliding system.

The new measurement also allowed for a more accurate determination of the 
$\pt$-integrated ${\rm c\overline{c}}$ production cross section at mid-rapidity
in pp collisions at $\sqrt{s}=7~\TeV$:
\begin{equation}
\nonumber
\left. {\rm d}\sigma^{\rm c\overline{c}}_{\rm pp,\,7\,TeV}/{\rm d}y \, \right|_{|y|<0.5}
\, = \,  954 \pm 69 \,({\rm stat})\, \pm 97 \,({\rm tot.~syst.}) ~\mu{\rm b}\,.
\end{equation}
In particular, the total systematic uncertainty of this measurement is about $\pm10\%$, while it was $^{+13}_{-21}\%$ for the previously-published measurement~\cite{Adam:2016ich}.

The total ${\rm c\overline{c}}$ production cross section in full phase space was calculated by combining the above measurement at mid-rapidity with 
that at forward rapidity by the LHCb Collaboration: 
\begin{equation}
\nonumber
{\rm \sigma^{\rm c\overline{c}}_{\rm pp,\,7\,TeV}({\rm ALICE,LHCb})= 7.44 \pm 0.14 \,({\rm stat.}) \, \pm 0.58 \,({\rm tot.~syst.})~{\rm mb}\,.}
\end{equation}

\newenvironment{acknowledgement}{\relax}{\relax}
\begin{acknowledgement}
\section*{Acknowledgements}

The ALICE Collaboration would like to thank all its engineers and technicians for their invaluable contributions to the construction of the experiment and the CERN accelerator teams for the outstanding performance of the LHC complex.
The ALICE Collaboration gratefully acknowledges the resources and support provided by all Grid centres and the Worldwide LHC Computing Grid (WLCG) collaboration.
The ALICE Collaboration acknowledges the following funding agencies for their support in building and running the ALICE detector:
A. I. Alikhanyan National Science Laboratory (Yerevan Physics Institute) Foundation (ANSL), State Committee of Science and World Federation of Scientists (WFS), Armenia;
Austrian Academy of Sciences and Nationalstiftung f\"{u}r Forschung, Technologie und Entwicklung, Austria;
Ministry of Communications and High Technologies, National Nuclear Research Center, Azerbaijan;
Conselho Nacional de Desenvolvimento Cient\'{\i}fico e Tecnol\'{o}gico (CNPq), Universidade Federal do Rio Grande do Sul (UFRGS), Financiadora de Estudos e Projetos (Finep) and Funda\c{c}\~{a}o de Amparo \`{a} Pesquisa do Estado de S\~{a}o Paulo (FAPESP), Brazil;
Ministry of Science \& Technology of China (MSTC), National Natural Science Foundation of China (NSFC) and Ministry of Education of China (MOEC) , China;
Ministry of Science, Education and Sport and Croatian Science Foundation, Croatia;
Ministry of Education, Youth and Sports of the Czech Republic, Czech Republic;
The Danish Council for Independent Research | Natural Sciences, the Carlsberg Foundation and Danish National Research Foundation (DNRF), Denmark;
Helsinki Institute of Physics (HIP), Finland;
Commissariat \`{a} l'Energie Atomique (CEA) and Institut National de Physique Nucl\'{e}aire et de Physique des Particules (IN2P3) and Centre National de la Recherche Scientifique (CNRS), France;
Bundesministerium f\"{u}r Bildung, Wissenschaft, Forschung und Technologie (BMBF) and GSI Helmholtzzentrum f\"{u}r Schwerionenforschung GmbH, Germany;
Ministry of Education, Research and Religious Affairs, Greece;
National Research, Development and Innovation Office, Hungary;
Department of Atomic Energy Government of India (DAE) and Council of Scientific and Industrial Research (CSIR), New Delhi, India;
Indonesian Institute of Science, Indonesia;
Centro Fermi - Museo Storico della Fisica e Centro Studi e Ricerche Enrico Fermi and Istituto Nazionale di Fisica Nucleare (INFN), Italy;
Institute for Innovative Science and Technology , Nagasaki Institute of Applied Science (IIST), Japan Society for the Promotion of Science (JSPS) KAKENHI and Japanese Ministry of Education, Culture, Sports, Science and Technology (MEXT), Japan;
Consejo Nacional de Ciencia (CONACYT) y Tecnolog\'{i}a, through Fondo de Cooperaci\'{o}n Internacional en Ciencia y Tecnolog\'{i}a (FONCICYT) and Direcci\'{o}n General de Asuntos del Personal Academico (DGAPA), Mexico;
Nationaal instituut voor subatomaire fysica (Nikhef), Netherlands;
The Research Council of Norway, Norway;
Commission on Science and Technology for Sustainable Development in the South (COMSATS), Pakistan;
Pontificia Universidad Cat\'{o}lica del Per\'{u}, Peru;
Ministry of Science and Higher Education and National Science Centre, Poland;
Korea Institute of Science and Technology Information and National Research Foundation of Korea (NRF), Republic of Korea;
Ministry of Education and Scientific Research, Institute of Atomic Physics and Romanian National Agency for Science, Technology and Innovation, Romania;
Joint Institute for Nuclear Research (JINR), Ministry of Education and Science of the Russian Federation and National Research Centre Kurchatov Institute, Russia;
Ministry of Education, Science, Research and Sport of the Slovak Republic, Slovakia;
National Research Foundation of South Africa, South Africa;
Centro de Aplicaciones Tecnol\'{o}gicas y Desarrollo Nuclear (CEADEN), Cubaenerg\'{\i}a, Cuba, Ministerio de Ciencia e Innovacion and Centro de Investigaciones Energ\'{e}ticas, Medioambientales y Tecnol\'{o}gicas (CIEMAT), Spain;
Swedish Research Council (VR) and Knut \& Alice Wallenberg Foundation (KAW), Sweden;
European Organization for Nuclear Research, Switzerland;
National Science and Technology Development Agency (NSDTA), Suranaree University of Technology (SUT) and Office of the Higher Education Commission under NRU project of Thailand, Thailand;
Turkish Atomic Energy Agency (TAEK), Turkey;
National Academy of  Sciences of Ukraine, Ukraine;
Science and Technology Facilities Council (STFC), United Kingdom;
National Science Foundation of the United States of America (NSF) and United States Department of Energy, Office of Nuclear Physics (DOE NP), United States of America.
\end{acknowledgement}

\bibliographystyle{utphys}
\bibliography{biblio}
%

%

\newpage
\appendix
\section{The ALICE Collaboration}
\label{app:collab}



\begingroup
\small
\begin{flushleft}
S.~Acharya$^\textrm{\scriptsize 139}$,
D.~Adamov\'{a}$^\textrm{\scriptsize 87}$,
M.M.~Aggarwal$^\textrm{\scriptsize 91}$,
G.~Aglieri Rinella$^\textrm{\scriptsize 34}$,
M.~Agnello$^\textrm{\scriptsize 30}$,
N.~Agrawal$^\textrm{\scriptsize 47}$,
Z.~Ahammed$^\textrm{\scriptsize 139}$,
N.~Ahmad$^\textrm{\scriptsize 17}$,
S.U.~Ahn$^\textrm{\scriptsize 69}$,
S.~Aiola$^\textrm{\scriptsize 143}$
% ...
\renewcommand\labelenumi{\textsuperscript{\theenumi}~}

\section*{Affiliation notes}
\renewcommand\theenumi{\roman{enumi}}
\begin{Authlist}
\item \Adef{0}Deceased
\item \Adef{idp1791728}{Also at: Dipartimento DET del Politecnico di Torino, Turin, Italy}
\item \Adef{idp1811120}{Also at: Georgia State University, Atlanta, Georgia, United States}
\item \Adef{idp3246128}{Also at: Also at Department of Applied Physics, Aligarh Muslim University, Aligarh, India}
\item \Adef{idp4017808}{Also at: M.V. Lomonosov Moscow State University, D.V. Skobeltsyn Institute of Nuclear, Physics, Moscow, Russia}
\end{Authlist}

\section*{Collaboration Institutes}
\renewcommand\theenumi{\arabic{enumi}~}

$^{1}$A.I. Alikhanyan National Science Laboratory (Yerevan Physics Institute) Foundation, Yerevan, Armenia
\\
$^{2}$Benem\'{e}rita Universidad Aut\'{o}noma de Puebla, Puebla, Mexico
\\
$^{3}$Bogolyubov Institute for Theoretical Physics, Kiev, Ukraine
\\
$^{4}$Bose Institute, Department of Physics 
and Centre for Astroparticle Physics and Space Science (CAPSS), Kolkata, India
\\
$^{5}$Budker Institute for Nuclear Physics, Novosibirsk, Russia
\\
$^{6}$California Polytechnic State University, San Luis Obispo, California, United States
\\
$^{7}$Central China Normal University, Wuhan, China
\\
$^{8}$Centre de Calcul de l'IN2P3, Villeurbanne, Lyon, France
\\
$^{9}$Centro de Aplicaciones Tecnol\'{o}gicas y Desarrollo Nuclear (CEADEN), Havana, Cuba
\\
$^{10}$Centro de Investigaciones Energ\'{e}ticas Medioambientales y Tecnol\'{o}gicas (CIEMAT), Madrid, Spain
\\
$^{11}$Centro de Investigaci\'{o}n y de Estudios Avanzados (CINVESTAV), Mexico City and M\'{e}rida, Mexico
\\
$^{12}$Centro Fermi - Museo Storico della Fisica e Centro Studi e Ricerche ``Enrico Fermi', Rome, Italy
\\
$^{13}$Chicago State University, Chicago, Illinois, United States
\\
$^{14}$China Institute of Atomic Energy, Beijing, China
\\
$^{15}$COMSATS Institute of Information Technology (CIIT), Islamabad, Pakistan
\\
$^{16}$Departamento de F\'{\i}sica de Part\'{\i}culas and IGFAE, Universidad de Santiago de Compostela, Santiago de Compostela, Spain
\\
$^{17}$Department of Physics, Aligarh Muslim University, Aligarh, India
\\
$^{18}$Department of Physics, Ohio State University, Columbus, Ohio, United States
\\
$^{19}$Department of Physics, Sejong University, Seoul, South Korea
\\
$^{20}$Department of Physics, University of Oslo, Oslo, Norway
\\
$^{21}$Department of Physics and Technology, University of Bergen, Bergen, Norway
\\
$^{22}$Dipartimento di Fisica dell'Universit\`{a} 'La Sapienza'
and Sezione INFN, Rome, Italy
\\
$^{23}$Dipartimento di Fisica dell'Universit\`{a}
and Sezione INFN, Cagliari, Italy
\\
$^{24}$Dipartimento di Fisica dell'Universit\`{a}
and Sezione INFN, Trieste, Italy
\\
$^{25}$Dipartimento di Fisica dell'Universit\`{a}
and Sezione INFN, Turin, Italy
\\
$^{26}$Dipartimento di Fisica e Astronomia dell'Universit\`{a}
and Sezione INFN, Bologna, Italy
\\
$^{27}$Dipartimento di Fisica e Astronomia dell'Universit\`{a}
and Sezione INFN, Catania, Italy
\\
$^{28}$Dipartimento di Fisica e Astronomia dell'Universit\`{a}
and Sezione INFN, Padova, Italy
\\
$^{29}$Dipartimento di Fisica `E.R.~Caianiello' dell'Universit\`{a}
and Gruppo Collegato INFN, Salerno, Italy
\\
$^{30}$Dipartimento DISAT del Politecnico and Sezione INFN, Turin, Italy
\\
$^{31}$Dipartimento di Scienze e Innovazione Tecnologica dell'Universit\`{a} del Piemonte Orientale and INFN Sezione di Torino, Alessandria, Italy
\\
$^{32}$Dipartimento Interateneo di Fisica `M.~Merlin'
and Sezione INFN, Bari, Italy
\\
$^{33}$Division of Experimental High Energy Physics, University of Lund, Lund, Sweden
\\
$^{34}$European Organization for Nuclear Research (CERN), Geneva, Switzerland
\\
$^{35}$Excellence Cluster Universe, Technische Universit\"{a}t M\"{u}nchen, Munich, Germany
\\
$^{36}$Faculty of Engineering, Bergen University College, Bergen, Norway
\\
$^{37}$Faculty of Mathematics, Physics and Informatics, Comenius University, Bratislava, Slovakia
\\
$^{38}$Faculty of Nuclear Sciences and Physical Engineering, Czech Technical University in Prague, Prague, Czech Republic
\\
$^{39}$Faculty of Science, P.J.~\v{S}af\'{a}rik University, Ko\v{s}ice, Slovakia
\\
$^{40}$Faculty of Technology, Buskerud and Vestfold University College, Tonsberg, Norway
\\
$^{41}$Frankfurt Institute for Advanced Studies, Johann Wolfgang Goethe-Universit\"{a}t Frankfurt, Frankfurt, Germany
\\
$^{42}$Gangneung-Wonju National University, Gangneung, South Korea
\\
$^{43}$Gauhati University, Department of Physics, Guwahati, India
\\
$^{44}$Helmholtz-Institut f\"{u}r Strahlen- und Kernphysik, Rheinische Friedrich-Wilhelms-Universit\"{a}t Bonn, Bonn, Germany
\\
$^{45}$Helsinki Institute of Physics (HIP), Helsinki, Finland
\\
$^{46}$Hiroshima University, Hiroshima, Japan
\\
$^{47}$Indian Institute of Technology Bombay (IIT), Mumbai, India
\\
$^{48}$Indian Institute of Technology Indore, Indore, India
\\
$^{49}$Indonesian Institute of Sciences, Jakarta, Indonesia
\\
$^{50}$Inha University, Incheon, South Korea
\\
$^{51}$Institut de Physique Nucl\'eaire d'Orsay (IPNO), Universit\'e Paris-Sud, CNRS-IN2P3, Orsay, France
\\
$^{52}$Institute for Nuclear Research, Academy of Sciences, Moscow, Russia
\\
$^{53}$Institute for Subatomic Physics of Utrecht University, Utrecht, Netherlands
\\
$^{54}$Institute for Theoretical and Experimental Physics, Moscow, Russia
\\
$^{55}$Institute of Experimental Physics, Slovak Academy of Sciences, Ko\v{s}ice, Slovakia
\\
$^{56}$Institute of Physics, Academy of Sciences of the Czech Republic, Prague, Czech Republic
\\
$^{57}$Institute of Physics, Bhubaneswar, India
\\
$^{58}$Institute of Space Science (ISS), Bucharest, Romania
\\
$^{59}$Institut f\"{u}r Informatik, Johann Wolfgang Goethe-Universit\"{a}t Frankfurt, Frankfurt, Germany
\\
$^{60}$Institut f\"{u}r Kernphysik, Johann Wolfgang Goethe-Universit\"{a}t Frankfurt, Frankfurt, Germany
\\
$^{61}$Institut f\"{u}r Kernphysik, Westf\"{a}lische Wilhelms-Universit\"{a}t M\"{u}nster, M\"{u}nster, Germany
\\
$^{62}$Instituto de Ciencias Nucleares, Universidad Nacional Aut\'{o}noma de M\'{e}xico, Mexico City, Mexico
\\
$^{63}$Instituto de F\'{i}sica, Universidade Federal do Rio Grande do Sul (UFRGS), Porto Alegre, Brazil
\\
$^{64}$Instituto de F\'{\i}sica, Universidad Nacional Aut\'{o}noma de M\'{e}xico, Mexico City, Mexico
\\
$^{65}$IRFU, CEA, Universit\'{e} Paris-Saclay, F-91191 Gif-sur-Yvette, France, Saclay, France
\\
$^{66}$iThemba LABS, National Research Foundation, Somerset West, South Africa
\\
$^{67}$Joint Institute for Nuclear Research (JINR), Dubna, Russia
\\
$^{68}$Konkuk University, Seoul, South Korea
\\
$^{69}$Korea Institute of Science and Technology Information, Daejeon, South Korea
\\
$^{70}$KTO Karatay University, Konya, Turkey
\\
$^{71}$Laboratoire de Physique Corpusculaire (LPC), Clermont Universit\'{e}, Universit\'{e} Blaise Pascal, CNRS--IN2P3, Clermont-Ferrand, France
\\
$^{72}$Laboratoire de Physique Subatomique et de Cosmologie, Universit\'{e} Grenoble-Alpes, CNRS-IN2P3, Grenoble, France
\\
$^{73}$Laboratori Nazionali di Frascati, INFN, Frascati, Italy
\\
$^{74}$Laboratori Nazionali di Legnaro, INFN, Legnaro, Italy
\\
$^{75}$Lawrence Berkeley National Laboratory, Berkeley, California, United States
\\
$^{76}$Moscow Engineering Physics Institute, Moscow, Russia
\\
$^{77}$Nagasaki Institute of Applied Science, Nagasaki, Japan
\\
$^{78}$National and Kapodistrian University of Athens, Physics Department, Athens, Greece, Athens, Greece
\\
$^{79}$National Centre for Nuclear Studies, Warsaw, Poland
\\
$^{80}$National Institute for Physics and Nuclear Engineering, Bucharest, Romania
\\
$^{81}$National Institute of Science Education and Research, Bhubaneswar, India
\\
$^{82}$National Nuclear Research Center, Baku, Azerbaijan
\\
$^{83}$National Research Centre Kurchatov Institute, Moscow, Russia
\\
$^{84}$Niels Bohr Institute, University of Copenhagen, Copenhagen, Denmark
\\
$^{85}$Nikhef, Nationaal instituut voor subatomaire fysica, Amsterdam, Netherlands
\\
$^{86}$Nuclear Physics Group, STFC Daresbury Laboratory, Daresbury, United Kingdom
\\
$^{87}$Nuclear Physics Institute, Academy of Sciences of the Czech Republic, \v{R}e\v{z} u Prahy, Czech Republic
\\
$^{88}$Oak Ridge National Laboratory, Oak Ridge, Tennessee, United States
\\
$^{89}$Petersburg Nuclear Physics Institute, Gatchina, Russia
\\
$^{90}$Physics Department, Creighton University, Omaha, Nebraska, United States
\\
$^{91}$Physics Department, Panjab University, Chandigarh, India
\\
$^{92}$Physics Department, University of Cape Town, Cape Town, South Africa
\\
$^{93}$Physics Department, University of Jammu, Jammu, India
\\
$^{94}$Physics Department, University of Rajasthan, Jaipur, India
\\
$^{95}$Physikalisches Institut, Eberhard Karls Universit\"{a}t T\"{u}bingen, T\"{u}bingen, Germany
\\
$^{96}$Physikalisches Institut, Ruprecht-Karls-Universit\"{a}t Heidelberg, Heidelberg, Germany
\\
$^{97}$Physik Department, Technische Universit\"{a}t M\"{u}nchen, Munich, Germany
\\
$^{98}$Purdue University, West Lafayette, Indiana, United States
\\
$^{99}$Pusan National University, Pusan, South Korea
\\
$^{100}$Research Division and ExtreMe Matter Institute EMMI, GSI Helmholtzzentrum f\"ur Schwerionenforschung GmbH, Darmstadt, Germany
\\
$^{101}$Rudjer Bo\v{s}kovi\'{c} Institute, Zagreb, Croatia
\\
$^{102}$Russian Federal Nuclear Center (VNIIEF), Sarov, Russia
\\
$^{103}$Saha Institute of Nuclear Physics, Kolkata, India
\\
$^{104}$School of Physics and Astronomy, University of Birmingham, Birmingham, United Kingdom
\\
$^{105}$Secci\'{o}n F\'{\i}sica, Departamento de Ciencias, Pontificia Universidad Cat\'{o}lica del Per\'{u}, Lima, Peru
\\
$^{106}$Sezione INFN, Bari, Italy
\\
$^{107}$Sezione INFN, Bologna, Italy
\\
$^{108}$Sezione INFN, Cagliari, Italy
\\
$^{109}$Sezione INFN, Catania, Italy
\\
$^{110}$Sezione INFN, Padova, Italy
\\
$^{111}$Sezione INFN, Rome, Italy
\\
$^{112}$Sezione INFN, Trieste, Italy
\\
$^{113}$Sezione INFN, Turin, Italy
\\
$^{114}$SSC IHEP of NRC Kurchatov institute, Protvino, Russia
\\
$^{115}$Stefan Meyer Institut f\"{u}r Subatomare Physik (SMI), Vienna, Austria
\\
$^{116}$SUBATECH, IMT Atlantique, Universit\'{e} de Nantes, CNRS-IN2P3, Nantes, France
\\
$^{117}$Suranaree University of Technology, Nakhon Ratchasima, Thailand
\\
$^{118}$Technical University of Ko\v{s}ice, Ko\v{s}ice, Slovakia
\\
$^{119}$Technical University of Split FESB, Split, Croatia
\\
$^{120}$The Henryk Niewodniczanski Institute of Nuclear Physics, Polish Academy of Sciences, Cracow, Poland
\\
$^{121}$The University of Texas at Austin, Physics Department, Austin, Texas, United States
\\
$^{122}$Universidad Aut\'{o}noma de Sinaloa, Culiac\'{a}n, Mexico
\\
$^{123}$Universidade de S\~{a}o Paulo (USP), S\~{a}o Paulo, Brazil
\\
$^{124}$Universidade Estadual de Campinas (UNICAMP), Campinas, Brazil
\\
$^{125}$Universidade Federal do ABC, Santo Andre, Brazil
\\
$^{126}$University of Houston, Houston, Texas, United States
\\
$^{127}$University of Jyv\"{a}skyl\"{a}, Jyv\"{a}skyl\"{a}, Finland
\\
$^{128}$University of Liverpool, Liverpool, United Kingdom
\\
$^{129}$University of Tennessee, Knoxville, Tennessee, United States
\\
$^{130}$University of the Witwatersrand, Johannesburg, South Africa
\\
$^{131}$University of Tokyo, Tokyo, Japan
\\
$^{132}$University of Tsukuba, Tsukuba, Japan
\\
$^{133}$University of Zagreb, Zagreb, Croatia
\\
$^{134}$Universit\'{e} de Lyon, Universit\'{e} Lyon 1, CNRS/IN2P3, IPN-Lyon, Villeurbanne, Lyon, France
\\
$^{135}$Universit\'{e} de Strasbourg, CNRS, IPHC UMR 7178, F-67000 Strasbourg, France, Strasbourg, France
\\
$^{136}$Universit\`{a} degli Studi di Pavia, Pavia, Italy
\\
$^{137}$Universit\`{a} di Brescia, Brescia, Italy
\\
$^{138}$V.~Fock Institute for Physics, St. Petersburg State University, St. Petersburg, Russia
\\
$^{139}$Variable Energy Cyclotron Centre, Kolkata, India
\\
$^{140}$Warsaw University of Technology, Warsaw, Poland
\\
$^{141}$Wayne State University, Detroit, Michigan, United States
\\
$^{142}$Wigner Research Centre for Physics, Hungarian Academy of Sciences, Budapest, Hungary
\\
$^{143}$Yale University, New Haven, Connecticut, United States
\\
$^{144}$Yonsei University, Seoul, South Korea
\\
$^{145}$Zentrum f\"{u}r Technologietransfer und Telekommunikation (ZTT), Fachhochschule Worms, Worms, Germany
\endgroup

\end{document}